\documentclass[aps,prb,twocolumn,float,amsmath]{revtex4}
\usepackage[dvips]{color}
\usepackage[normalem]{ulem} 
 \usepackage{hyperref}
\definecolor{g-blue}{rgb}{0.83,0.95,1}
\definecolor{g-yellow}{rgb}{1,1,0.7}
\definecolor{g-green}{rgb}{0.9,1,0.9}
\definecolor{green}{rgb}{0,0.6,0}
\definecolor{cyan}{rgb}{0,0.7,0.7}
\definecolor{black}{rgb}{0,0,0}
\definecolor{grey}{rgb}{0.4 ,0.4 ,0.4 }

\usepackage{bm}
\usepackage{amssymb}
\usepackage{amsfonts}
\usepackage{amsmath}
 \usepackage{graphicx}
  \usepackage{amsmath,bm,epsfig}
\def \ed {\end{document}}
\def\Fbox#1{\vskip1ex\hbox to 8.5cm{\hfil\fboxsep0.3cm\fbox{%
  \parbox{8.0cm}{#1}}\hfil}\vskip1ex\noindent}  

\newcommand{\eq}[1]{(\ref{#1})}
\newcommand{\Eq}[1]{Eq.\,(\ref{#1})}
\newcommand{\Eqs}[1]{Eqs.\,(\ref{#1})}
\newcommand{\Fig}[1]{Fig.\,\ref{#1}}
\newcommand{\Figs}[1]{Figs.\,\ref{#1}}
\newcommand{\Sec}[1]{Sec.\,\ref{#1}}
\newcommand{\Ref}[1]{Ref.\,\cite{#1}}
\newcommand{\Refs}[1]{Refs.\,\cite{#1}}

\def\be{\begin{equation}}\def\ee{\end{equation}}
\def\bea{\begin{eqnarray}}\def\eea{\end{eqnarray}}
\def\bse{\begin{subequations}}\def\ese{\end{subequations}}
\newcommand{\BE}[1]{\begin{equation}\label{#1}}
\newcommand{\BEA}[1]{\begin{eqnarray}\label{#1}}
\newcommand{\BSE}[1]{\begin{subequations}\label{#1}}

\let \nn  \nonumber  \newcommand{\br}{\\ \nn}

\let \= \equiv \let\*\cdot \let\~\widetilde \let\^\widehat \let\-\overline
\let\p\partial

  \def\1{\bm1} 

\def\<{\left\langle}    \def\>{\right\rangle}
\def\({\left(}          \def\){\right)}
 \def \[ {\left [} \def \] {\right ]}

\renewcommand{\a}{\alpha}\newcommand{\g}{\gamma}
\newcommand{\G} {\Gamma}\renewcommand{\d}{\delta}
\newcommand{\D}{\Delta}\newcommand{\ve}{\varepsilon}
\renewcommand{\o}{\omega} 
\renewcommand{\L}{\Lambda}

\def\r{\rho}\def\k{\kappa}

\newcommand{\B}[1]{{\bm{#1}}}
\newcommand{\C}[1]{{\mathcal{#1}}}    

\renewcommand{\sb}[1]{_{\text {#1}}}  
\renewcommand{\sp}[1]{^{\text {#1}}}  
\newcommand{\Sp}[1]{^{^{\text {#1}}}} 
\def\Sb#1{_{\scriptscriptstyle\rm{#1}}}

\def\He4 {$^4$He~}

\begin{document}

\title{Energy and Vorticity Spectra in Turbulent Superfluid $^4$He from $T=0$ to $T_\lambda$.}
\author{Laurent Bou\'e$^*$,  Victor S. L'vov$^*$, Yotam Nagar$^*$, Sergey V. Nazarenko$^\dag$, Anna Pomyalov$^*$, and Itamar
Procaccia$^*$  }
\affiliation{$^*$Department of Chemical Physics,  Weizmann Institute  of Science, Rehovot 76100, Israel \\
$^\dag$University of Warwick, Mathematics Institute Coventry, CV4 7AL, UK}

\begin{abstract}
We discuss the energy and vorticity spectra of turbulent superfluid $^4$He in the entire temperature range from $T=0$ up
to the phase transition  ``$\lambda$ point", $T_\lambda\simeq 2.17\,$K. Contrary to classical developed turbulence
in which there are only two typical scales, i.e. the energy injection $L$ and the dissipation scales $\eta$, here the quantization
of vorticity introduces two additional scales,  the vortex core radius $a_0$ and the mean vortex spacing $\ell$.
We present these spectra for the super- and the  normal-fluid components in the entire range of scales from $L$ to $a_0$
including the cross-over scale $\ell$ where the hydrodynamic eddy-cascade is replaced by the cascade of Kelvin
 waves on individual vortices. At this scale a bottleneck accumulation of the energy was found earlier at $T=0$.
 We show that  even very small mutual friction dramatically suppresses the bottleneck effect due to the dissipation of the Kelvin
 waves. Using our results for the spectra we estimate the Vinen ``effective viscosity" $\nu'$ in the entire temperature range and
 show agreement with numerous experimental observation for $\nu'(T)$.
\end{abstract}

\maketitle

\section{\label{s:intro} Introduction}

 Superfluidity was discovered  by Kapitza and by Allen and
Misener in 1938 who  demonstrated the existence of an inviscid fluid flow of $^4$He  below $T_\lambda\simeq 2.17\,$K.  In the same year, London linked the properties of  the superfluid \He4 to Bose-Einstein condensation.

Soon after, Landau and Tisza offered a ``two fluid" model
in which the dynamics of the superfluid $^4$He is described in terms of a viscous normal
component and an inviscid superfluid component, each with its own density $\rho\sb n(T)$ and $\rho\sb s(T)$ and its own velocity field $\B u\sb n(\B r,t)$ and $\B u\sb s(\B r,t)$.  Already in 1955, Feynman realized\,\cite{Feynman} that the potential appearance of quantized
vortex lines will result in a new type of turbulence, the turbulence of superfluids. The experimental verification of this
prediction followed in the paper by Hall and Vinen a year later\,\cite{Vinen}.

An isolated vortex line is  a stable topological defect in which the superfluid density drops to zero and the velocity $v\sb s \propto 1/r$. Here $r> a_0$ is the radial distance from the center that exceeds a core radius $a_0\simeq 10^{-8}\,$cm in $^4$He.
The existence of quantized vortex lines\cite{1,2,Vinen2008} in superfluid tubulence introduces automatically additional length scales that do not exist in classical turbulence. In addition to $a_0$, the density of vortex lines $\C L$ defines an ``inter-vortex" average spacing denoted as $\ell\equiv 1/\sqrt{\C L}$.

The pioneering experimental observation of Maurer and Tabeling\,\cite{MT-98}
  showed quite clearly that the {\em large-scale} energy spectrum of turbulent $^4$He
above and below $T_\lambda$ are indistinguishable. This and later experiments and simulations gave rise to the growing consensus that on scales much larger than $\ell$ the energy spectra of turbulent superfluids are very close to those of classical fluids if they are excited similarly\,\cite{BLR}.  The understanding is
that motions with scales $R \gg \ell$ correlate the vortex
lines, organizing them into vortex bundles. At these large scales the quantization of the vortex lines becomes irrelevant and the large scale  motions are similar to those of continuous hydrodynamic eddies. Obviously, since energy is cascaded by hydrodynamic eddies to smaller and smaller scales, we must reach a scale where the absence of viscous dissipation will require new physics.

Evidently, when the observation scales approach $\ell$ and below, the discreteness of the quantized vortex lines becomes
crucial. Indeed, on such scales the dynamics of the vortex lines themselves become relevant including vortex reconnections and the
excitation of Kelvin waves on the individual vortex lines. Kelvin waves exist also in classical hydrodynamics but here they
become important in taking over the role of transferring the energy further down the scales. Their nonlinear interaction
results in a so-called ``weak wave turbulence"\,\cite{92ZLF,11Nazar} supporting a mean energy flux towards shorter and shorter scales. Finally when
the cascade reaches the core radius scale the energy is radiated away by quasiparticles (phonons in $^4$He )\,\cite{VinenNiemela}.

Although the overall picture of superfluid turbulence described above seems quite reasonable, some important details
are yet to be established. A particularly interesting issue is the physics on scales close to the crossover between the eddy-dominated and the Kelvin wave-dominated regimes of the spectrum. It was pointed out in \Ref{LNR-1} that the nonlinear transfer mechanism of weakly
nonlinear Kelvin waves on sparse vortex lines is less efficient than the energy transfer mechanism due to strongly nonlinear eddy
interactions in continuous fluids. This may cause an energy cascade stagnation at the crossover scale.

The present paper is motivated by some exciting new experimental and simulational developments \cite{Lathrop1,Lathrop2,Andreev,Ladik,Ladik1,Ladik2,Roche,Manchester-exp,Golov2,Karman,rev4,exp,Helsinki-exp,Tsubota2008,araki,stalp,nimiela2005,exp1,Vin03-PRL,cn,exp2,exp3,exp4,RocheBarenghi} that call for a fresh analysis
of the physics of superfluid turbulence in a range of temperatures and length scales. These developments include, among others, cryogenic flow visualization techniques by micron-sized solid particles and metastable helium molecules, that allow, e.g. direct observation of vortex reconnections, mean normal and superfluid velocity profiles in thermal counterflow \,\cite{Lathrop1,Lathrop2}; the observation of Andreev reflection by an array of vibrating wire sensors shedding light on the role of vortex dynamics in the formation of quantum turbulence, etc.\,\cite{Andreev}; the measurements of the vortex line density by the attenuation of second sound \cite{Ladik,Ladik1,Ladik2,Roche} and by the
attenuation of  ion beams \cite{Manchester-exp,Golov2}.
 An important role in the recent developments is played by large-scale well-controlled apparata, like  the Prague \cite{Ladik,Ladik1,Ladik2} and  the Grenoble wind tunnels\cite{Roche} , the Manchester spin-down \cite{Manchester-exp,Golov2}, the Grenoble Von Karman flows \cite{Karman},  the Helsinki rotating cryostat \cite{rev4,exp,Helsinki-exp}, and some other experiments. Additional insight was provided by  large-scale numerical simulations of quantum turbulence by the vortex-filament and other methods that gives direct access to detailed picture of vortex dynamics which is still unavailable in experiments, see also
~Refs.\cite{Tsubota2008,araki,Vin03-PRL,cn,RocheBarenghi,exp3}.

The stagnation of the energy cascade at the intervortex scale mentioned above is referred to as the bottleneck effect.
This issue was studied in \Ref{LNR-1} in the approximation of a ``sharp" crossover. \Ref{LNR-2} introduced a model of a gradual eddy-wave crossover in which both  the
eddy and the wave contributions to the energy spectrum of superfluid turbulence at zero temperature  $\C E\sp s(k)$ (see \Eq{tot-s})
were found as a continuous function of the wave vector $k$. The main message of
\Ref{LNR-2} is that the  \emph{bottleneck} phenomenon is robust and common to all the situations where the energy cascade
experiences a continuous-to-discrete transition. The details of the particular mechanism of this transition are secondary. Indeed,
most discrete physical processes are less efficient than their continuous  counterparts~\footnote{It is interesting to make
comparison with turbulence of weakly nonlinear waves where the main energy transfer mechanism is due to wavenumber and frequency
resonances. In bounded volumes the set of wave modes is discrete and there are much less resonances between them than in the
continuous case. Thus the energy cascades between scales are significantly suppressed.}. On the other hand, particular mechanisms of
the continuous-to-discrete transition can obviously lead to different strengths of the bottleneck effect.

The main goal of the present paper is to develop a theory of  superfluid turbulence that analyzes the dynamics of turbulent
superfluid $^4$He and
computes its energy and vorticity spectra in the entire temperature range from $T\to 0$ up to the phase transition, ``$\lambda$
 point" $T_\lambda \simeq 2.17\,$K, and in the entire range of scales $R$, from the outer (energy-containing, or energy-injection)
 scale $L$ down to the core radius $a_0$. We put a particular focus  on the crossover scales $R\sim \ell$, where the bottleneck energy
 accumulation is expected\,\cite{LNR-1,LNR-2}.

 The main results of this paper are  presented in \Sec{s:II}.  { Its introductory  subsection,
  }
 \begin{description}
\item  {\ref{ss:intro-A} Basic   approximations and models, }

  {overviews the basic physical mechanisms,  which determine the behavior of superfluid turbulence and  describes the set of main approximations and models, used for their description. \\~\\
 The rest  of \Sec{s:II} is devoted to the following problems: }

 \item \ref{ss:intro-B}. Temperature dependence of the energy spectra and the bottleneck effect in turbulent $^4$He;
 \item \ref{ss:intro-C}. Temperature dependence of the vorticity spectra;
 \item \ref{ss:intro-D}. Correlations between normal and superfluid motions and the energy exchange between components;
 \item \ref{ss:intro-E}. Temperature dependence of the effective superfluid viscosity in $^4$He.
 \end{description}

 Clearly, the basic physics of the large scale motions,  differ from that of small scale motions. The same can be said about different regions of temperature: zero temperature limit, small, intermediate and large temperatures. It would be difficult to follow the full description of the physical picture of superfluid turbulence in all these regimes without clear understanding of the entire phenomenon as a whole.  Therefore in \Sec{ss:intro-A} we restricted ourselves to a panoramic overview of the main approximations and models, leaving detailed consideration of some important, but in some sense secondary issues, to the next  two sections of the paper (\Sec{s:EnSp} and \Sec{s:1-fluid}). These include the  analysis of the  range of validity of the basic
equations of motions, of the main approximations made in the derivation, and of the numerical procedures. These sections consist of  the following subsections:
\begin{description}
 \item \ref{ss:2fm}. Coarse-grained, two-fluid, gradually-damped Hall-Vinen-Bekarevich-Khalatnikov (HVBK) equations;
 \item \ref{ss:2fluid}. Two-fluid Sabra shell-model of turbulent $^4$He;
 \item \ref{ss:E-HD}. Differential approximations for the energy fluxes of the hydrodynamic and Kelvin wave motions;
 \item \ref{s:BB}. One-fluid differential model of the graduate eddy-wave crossover;

  \end{description}
In the final  Section \,\ref{s:Disc} we summarize our results on the temperature dependence of the energy spectra of the normal and
superfluid components in the entire region of scales. We demonstrate in Fig.\,\ref{fig5}
that the computed temperature dependence of the effective viscosity $\nu'(T)$ agrees qualitatively with the experimental data in the
entire temperature range.  We
consider this agreement as a strong evidence that our low-temperature, one fluid differential model and the high temperature coarse-grained gradually damped HVBK model capture the relevant basic physics of the turbulent behavior of $^4$He.

\section{\label{s:II} Underlaying  physics and  the results}

\subsection{\label{ss:intro-A}  {Basic   approximations and models}}
\subsubsection{\label{sss:2fm}  {Coarse-grained, two-fluid, gradually-damped HVBK equations}}

 As we noticed in the Introduction, the  large-scale motions of superfluid \He4 (with characteristic scales $R\gg \ell$) are described using the two-fluid  model as
interpenetrating motions of a normal and a superfluid component with densities $\rho\sb n(\B r, t,T)$, $\rho\sb s(\B r, t,T)$ and
velocities $\B u \sb n(\B r, t)$, $\B u\sb s(\B r, t)$.  Following
\Ref{L199}  we   neglect variations of densities by considering them as functions of the temperature $T$ only, $\rho\sb n(T)$ and $\rho\sb s(T)$. We also neglect both the bulk viscosity and  the thermal conductivity. This results in the simplest form of the two incompressible-fluids model for superfluid  $^4$He that have a form of the Euler equation for   $\B u\sb s$  and the
Navier-Stokes equation for   $\B u\sb n$, see   e.g. Eqs. (2.2) and
(2.3) in  Donnely's textbook~\cite{1}. As motivated below, we add an effective superfluid viscosity term also in the superfluid equation, writing
\begin{subequations}\label{NSE}
\begin{eqnarray}\label{NSEa} 
 \frac{\p \,\B u\sb s}{\p t}+ (\B u\sb s\* \B
\nabla)\B u\sb s   - \frac 1{\rho\sb s }\B \nabla p\sb s&=&\nu'\sb s\,  \Delta \B u\sb s -\B F \sb {ns}\,, 
 \\ \label{NSEb}
  \frac{\p \,\B u\sb n}{\p t}+ (\B u\sb n\* \B
\nabla)\B u\sb n  - \frac 1{\rho\sb n }\B \nabla p\sb n&=&\nu\sb n\,  \Delta \B
u\sb n + \frac{\rho\sb s}{\rho\sb n}\B F \sb {ns}\ . ~~~~~~~
\end{eqnarray} 
Here $p\sb n$,  $p\sb s$   are  the pressures  of the normal
and the superfluid components:
$$ p\sb n =\frac{\rho\sb n}{\rho }[p+\rho\sb s|\B u\sb s-\B u\sb
n|^2]\,,\
 p\sb s = \frac{\rho\sb s}{\rho }[p-\rho\sb n|\B u\sb s-\B u\sb
n|^2]\,,
  $$
$\rho\= \rho\sb s+\rho\sb n$  is the total density,     $\nu\sb n$ is the kinematic viscosity of normal fluid.

The term $\B F\sb {ns}$ describes the mutual friction  between the superfluid and the normal components mediated by quantized vortices, which transfer momentum from the superfluid to the normal subsystem and vice versa. Following Ref.~\cite{LNV}, we approximate  it  as follows: 
\begin{equation}\label{NSEf-a}
  \B F\sb {ns}\simeq \a\,  \rho\sb s \bar \omega\sb s(\B u\sb s -\B
u\sb n) \,, \end{equation}
where   $\bar \omega\sb s$ is the characteristic
superfluid vorticity.

The equations\,\eqref{NSE} are referred to as the
Hall-Vinen-Bekarevich-Khalatnikov (or HVBK) coarse-grained model.
The relevant parameters in these equations,  are the densities $\rho\sb s (T)$ and $\rho\sb n (T)$, the mutual friction parameters $\alpha(T)$ and
 the kinematic viscosity of the normal-fluid component $\nu\sb n(T)$ normalized by $\rho\sb n$.

 The original  HVBK model does not take into account the important process of vortex reconnection. In fact, vortex
  reconnections are responsible for the dissipation  of the superfluid motion due to mutual friction. This extra dissipation can be modeled as an effective superfluid viscosity $\nu'\sb s(T)$ as suggested in \Ref{VinenNiemela}:
  \begin{equation}\label{nus}
 \nu'\sb s (T)\approx \alpha \, \kappa\ .
 \end{equation}\end{subequations}
 We have added  a dissipative term proportional to $\nu'\sb s $  to the standard HVBK model and the resulting Eqs.\,\eqref{NSE} [discussed in more details in \Sec{s:EnSp}] will be referred to as the ``gradually damped HVBK model". We use this  name  to distinguish our model from the
  alternative ``truncated HVBK  model"  suggested in \Ref{22} which was recently used for
 for numerical analysis of  the effective viscosity $\nu'(T)$ in \Ref{Ladik2}. We  suspect  that the sharp truncation introduced in the latter model
 creates an artificial bottleneck effect that is removed in the gradually damped model. The difference in predictions between the
  models will be further discussed in \Sec{ss:truncated-vs-gd}.

 \subsubsection{ {\label{sss:Sabra1}Two-fluid Sabra shell-model of turbulent $^4$He}}
The gradually damped HVBK \Eqs{NSE} provide an adequate basis for our studies of the large-scale statistics of superfluid turbulence. However their mathematical analysis is very difficult because of the same reasons that make the the Navier-Stokes equations\cite{LP-Exact}
difficult. The interaction term  is much larger than the linear part of the equation (their ratio  is the Reynolds number, Re$\gg 1$), the nonlinear term is nonlocal both in the physical   and in the wave-vector $\B k$-space, the energy exchange between eddies of similar scales, that determines the statistics of turbulence, is masked by much larger kinematic effect of sweeping of small eddies by larger ones, etc.

Direct numerical simulations  of the HVBK \Eqs{NSE} are even more difficult than the Navier-Stokes analog, being extremely demanding computationally,  allowing  therefore for a very short span of scales. A  possible  simplification is provided by shell models of turbulence\,\cite{9,10,12,13,14,Sabra,S2,S3,S4,S5,Bif,s1,s2,s3,s4,s5}.  They   significantly simplify  the Navier-Stokes equations for space-homogeneous, isotropic turbulence of incompressible fluid. The idea is to  consider the equations in wave vector $\B k$-Fourier representation and to mimic   the  statistics  of $\B u(\B k,t)$ in the entire
shell of wave  numbers  $k_m < k_{m+1}$  by only one complex shell
velocity $v_m$. The integer index $m$ is referred to as the shell
index, and the shell wave numbers are chosen as a geometric
progression $k_m = k_0\lambda^m$, with $\lambda$ being the shell-spacing
parameter. This results in the ordinary differential equation
\begin{subequations}\label{sm}\begin{equation}\label{smA}
\Big( \frac{d}{d t}+ \nu k_m^2\Big ) v_m= \mbox{NL}_m\{v_{m'}\}\ .
\end{equation}
 Here the nonlinear term NL$_m\{v_m'\}$ is linear in $k$ and quadratic in $v_{m'}$ ( a functional of the  set $\{v_m'\}$), which usually involves shell velocities with $|m-m'|\leq 2$ .
 the kinetic energy is preserved by the nonlinear term. For example, in the popular  Gledzer-Ohkitani-Yamada
(GOY) shell model\cite{9,10}
\begin{eqnarray}\nonumber
\mbox{NL}_m\{v_{m'} \}&=&ik_m(a \lambda v_{m+2}v_{m+1}+ b v_{m-1}v_{m+1}\\ \label{GOY}
&& + c v_{m-1}v_{m-2})^*\,,\quad \mbox{GOY}\ ,
\end{eqnarray}\end{subequations}
where the asterisk $^*$ stands for complex conjugation. In the limit $\nu\to 0$ and with $a+b+c=0$, \Eqs{sm} preserve the kinetic energy $E=\sum_m |v_m|^2$ and has a second  integral of motion $H=\sum_m (a/c)^m|v_m|^2$. The traditional choice  $a=\lambda |b|$ allows to associate $H$ with the helicity in the Navier-Stokes equations.

 Note that the simultaneous rescaling   $a \Rightarrow a p $, $b \Rightarrow b p $ and  $c\Rightarrow c p $ with some factor  $p$ results
 in  a straightforward rescaling of the  the time variable $t \Rightarrow t/p$ without any effect on the  instantaneous  stationary statistics of the model.
 Thus, the shell model\,\eqref{sm} has only one fitting parameter $\lambda$, which has only little effect on the resulting statistics.  The  traditional choice $\lambda=2$ allows to reasonably model the  interactions  in $k$-space with  an efficient energy exchange between modes  of similar index $m$.

We stress that with the above choice of parameters, $a+b+c=0$, $a=\lambda |b|$, and
 \begin{equation}\label{choice}
 \lambda=2\,, \quad     a=1\,, \  b=c=-0.5\,,
  \end{equation}
  the shell models  reproduced  well various   scaling properties  of space-homogeneous, isotropic turbulence of incompressible fluids, see Ref.\cite{Bif} and references therein.  To mention just a few:

  -- the values of anomalous scaling exponents (see, e.g. Table I in\,\cite{Sabra});

   -- the viscous corrections to the scaling exponents\,\cite{S2};

  -- the connection between extreme events (outliers) and multiscaling\,\cite{S3};

  -- the inverse cascade in two-dimensional turbulence\,\cite{S4};

  -- the strong universality in forced and decaying turbulence,\cite{S5}, etc\,\cite{s1,s2,s3,s4,s5}.


  Therefore, we propose shell models are a possible alternative to the numerical solution of the HVBK \Eqs{NSE}. This option was studied in Ref. \cite{8} which proposed a two-fluid GOY shell model for superfluid turbulence with an additional coupling  by  the mutual friction.

In our studies of superfluid turbulence\,\cite{Sabra1,Sabra2,BLPP-2013} and below,  we use the so-called Sabra-shell model\,\cite{Sabra},  with  a different form of the  nonlinear term:
\begin{eqnarray}\label{Sabra}\nonumber
\mbox{NL}_m\{v_{m'} \}&=&ik_m(a \lambda v_{m+2}v_{m+1}^*+ b v_{m-1}^*v_{m+1}\\ \label{Sabra}
&& - c v_{m-1}v_{m-2})\,,\quad \mbox{Sabra}\ .
\end{eqnarray}
The advantage of the Sabra model   over  the GOY model is that  the resulting spectra do not suffer from the  unphysical  period-three oscillations, thanks to the strong statistical locality induced by the phase invariance\,\cite{Sabra,Bif}.

 We solved  numerically the  two-fluid  Sabra-shell  model form  of the HVBK  equations\,\eqref{smA} and \eqref{Sabra}  coupled by the mutual friction,  for the shell
 velocities.  Gathering enough statistics, we computed the pair- and cross-correlation functions of the normal- and the super-fluid shell velocities. This led to the  energy spectra $\C E\sb{n }(k)$ and $\C E\sb{s }(k)$ together with the cross-correlation  $\C E\sb{ns}(k)$.

In the simulations  we used $32$ shells.  All the results are obtained by averaging over about 500 large eddy turnover times. The rest of details of the numerical implementation and simulations are given in \Sec{ss:2fluid}.

\subsubsection{\label{sss:LT-1fluid}   { Low temperature one-fluid eddy-wave model  of superfluid turbulence}}

 As we just explained, in the high-temperature region the fluid motions with scales $R< \ell$ are damped and motions with $R>\ell$ are  faithfully described by the  Sabra-shell  model \,\eqref{smA} and \eqref{Sabra}. In this approach we first solve the dynamical equation and then perform the statistical averaging numerically.

  In the low temperature regime, $T \lesssim T_\lambda /2$, where the Kelvin wave motions of individual vortex lines are important this approach is no longer tenable. Instead,  we
adopt a different strategy, in which we first perform the statistical averaging analytically and then solve the resulting equations for the averaged quantities numerically.

To this end we begin with the dynamical Biot-Savart equation of motion for quantized vortex lines. Then we applied the Hamiltonian description to develop a ``weak turbulence" formalism to the energy cascade by Kelvin waves \cite{92ZLF}. This
approach results in a closed form expression for the Kelvin wave energy spectra, derived in Ref.~\onlinecite{LN-09}:
\begin{subequations}\label{LNspec}\be\label{LN-s}
\C E\Sb{KW} (k)  = \frac{ C\Sb {LN}}{\Psi^{2/3}}  \,   \frac{\L \, \k \, \ve\Sb{KW}^{1/3}}{   k^{5/3}\ell ^{4/3}} \,, \quad \mbox{LN-spectrum}.
\end{equation}
Here  $\ve\Sb{KW}$ is the energy flux over small-scale region, $R<\ell$, and  $\Lambda \simeq \ln (\ell/a_0)$.  The value of the universal constant $C\Sb{LN}\approx 0.304$ was estimated analytically  in Ref.~\onlinecite{KW-2}.  The dimensionless
constant  $\Psi$ may be considered as  the r.m.s.  vortex line  deflection angle   at scale $\ell$ and is  given by
   \begin{equation}\label{int-psi}
\Psi\=  \frac  {8\pi\, \C E\Sb{KW} \ell^2} {\Lambda\, \kappa^2} \ .
\end{equation}\end{subequations}

In the low-temperature region, $T \lesssim T\sb c/2$,  the density of the normal component is very small and due to very large kinematic viscosity it may be considered at rest. Therefore the large scale motions of $^4$He, $R>\ell$, are governed by the first of HVBK \Eq{NSEa}, which coincides with the Navier-Stokes equation in the limit $T\to 0$.
Therefore, in the hydrodynamic range of scales, $R> \ell$, we can use  the
Kolmogorov-Obukhov $5/3$--law \cite{Frisch} for the hydrodynamic energy spectrum:
\begin{equation}\label{V6-1} 
\C E\Sb{HD}(k)=C\Sb{K41} \ve\Sb{HD} ^{2/3} k^{-5/3}\,, \quad \mbox{KO-41}\ .
\end{equation}
 Here $\ve\Sb{HD}$ is the energy flux over large scale range  and   $C\Sb{K41}\sim 1$ is the Kolmogorov dimensionless constant.

 Both spectra,\ \eq{LNspec} and \eq{V6-1} have the same $k$-dependence, $\propto k^{-5/3}$, but different powers of the energy flux. A way to match these spectra in the  $T\to 0$ limit was suggested in \Ref{LNR-2}.  The idea was to adopt the differential approximations to the Kelvin-wave\,\cite{KW-T} and hydrodynamic-energy flux\,\cite{Leith67}, based on their spectra\,\eqref{LN-s} and \eqref{V6-1}:
 \begin{subequations}\begin{eqnarray}
  \label{Leith-n}
  \varepsilon\Sb{HD}(k) &=& -{1\over 8} \, \sqrt{k^{11} \C E\Sb{HD}(k)}\  {d\,  \over d k} \frac{\C E\Sb{HD}(k) }{  k^2}\,, \\ \label{vareps-KW}
    \varepsilon\Sb{KW}(k)&=&-\frac{3\,\C E\Sb{KW}^2(k)\Psi^2k^6\ell^4}{5 (C\Sb {LN}\Lambda \kappa)^3}\, \frac {d\,  \C E\Sb{KW} (k)}{d
    k}\,,
    \end{eqnarray}
and to construct a differential approximation for the superfluid energy flux $\varepsilon$
 that is valid for all wave numbers (including the cross over scale):
 \begin{equation}
 \varepsilon(k)= \varepsilon\Sb{HD}(k)+\varepsilon\Sb{KW}(k)+\varepsilon\Sb{HD}\Sp{KW}(k)
 +\varepsilon\Sb{KW}\Sp{HD}(k)\ .
 \end{equation}\end{subequations}
  The
  additional cross-contributions   $\ve\Sb{HD } \Sp{KW}(k)$  and
   $\ve\Sb{KW } \Sp{HD}(k)$  originate from the interaction  of  two types of motion, hydrodynamic and Kelvin waves.

For $T\to 0$ the total energy flux should be $k$-independent, $\varepsilon(k)=$const. As explained in \Ref{LNR-2} this leads to an ordinary differential equation for the total superfluid
energy $\C E\sb s(k)= \C E\Sb{HD}(k)+\C E\Sb{KW}(k)$. In this paper we generalize this approach to the full temperature
interval with the help of the   energy balance equation
\begin{equation}\label{sBal}
\frac{\p \C E\sb s (k,t)}{\p t}+ \frac{\p \varepsilon\sb s (k,t)}{\p k} = \nu'\sb s k^2 \C E \sb s (k,t)\ .
\end{equation}
The  right-hand-side of this equation originates from the Vinen-Niemella viscosity in \Eq{NSEa} and accounts for the dissipation in the system.

Much more detailed description  of this procedure can be found in \Sec{s:1-fluid}.

 \subsection{\label{ss:intro-B}  Temperature dependence of the energy spectra and the bottleneck effect in turbulent $^4$He}
\begin{table*}
\begin{tabular}{||c|  c|c|c|  c| c|c|c|c|c|c|c|c|c|c|c|c|c||}
\hline\hline
 $T$, K& 0.43& 0.55& 0.8& 0.9&1.0& 1.1& 1.2&1.3 & 1.4&1.5& 1.6&1.7&1.8  & 1.9 &2.0& 2.1  & 2.16 \\
 \hline\hline
 $\rho\sb n/\rho$& --  & --&$9.3\cdot 10^{-4}$&0.003&0.007&0.014&0.026& 0.045&0.0728 &0.111& 0.162 & 0.229&0.313& 0.420&0.553&0.741
 &0.907\\ \hline
$\alpha$&    $10^{-6}$&$10^{-5}$&$6.5\cdot 10^{-4}$&0.0025  &0.0056 & $0.011$ & 0.026 & 0.034    & 0.051 & 0.072 &0.097&0.126&0.160
&0.206 &0.279 &0.48 &  1.097  \\

   $\alpha\rho/\rho\sb n$ &--&--&0.70 &0.83 &0.80 &0.78 & 1.00 & 0.76& 0.70& 0.65&0.60&0.55 &0.51 &0.49  & 0.53&0.65 & 1.209\\
   \hline
  $  \nu /\kappa$ &-- &-- &1.09&0.43&0.27&0.17&0.12&0.10&0.10& 0.09&0.09&0.09&0.09 &0.093  &0.101 &0.124 & 0.154\\
  $  \nu\sb n/\kappa$     & -- & -- &  1179
&   148
&     38
&    11.1
&   4.62
&     2.34
&     1.32
&   0.84
&    0.56
&    0.39
&     0.29
&     0.22
&   0.182
&  0.167 &  0.170   \\
$\nu\sb s '/\kappa$ & --  &   -- &0.0067 &0.022&0.040 & 0.061 & 0.099  & 0.101 & 0.135 &0.171 & 0.207 &0.234& 0.237& 0.280& 0.312 &
0.427 &0.815 \\

 \hline\hline

\end{tabular}\caption{\label{t:1}The parameters of the superfluid $^4$He, taken from Refs.\,\cite{VinenNiemela,DonnelyBarenghi98}:
 the relative density of the normal component $\rho\sb n/\rho$,  the mutual friction parameter $\alpha$, the combination
 $\alpha\rho/\rho\sb n$ [which  weakly depends temperature and is responsible for the mutual friction density in  \Eq{NSE}],  He-II
 kinematic viscosity $\nu\equiv \mu/\rho $ ($\mu$ is the dynamic viscosity) and  the  kinematic viscosity of the normal-fluid
 component $\nu\sb n\equiv \mu/\rho\sb n$;
 the  effective superfluid viscosity $\nu'\sb s$ (inter- and extrapolation of Ref.\,\cite{VinenNiemela} results).   }
 \end{table*}


To discuss our results we define the energy spectrum $\C E(k)$ of isotropic turbulence (in one-dimensional  $k$-space) such that
 \begin{equation}\label{def0}
 \C E= \int\limits _0^\infty \C E(k)\, dk= \frac12 \< |\B u(\B r,t)|^2\>\,,
 \end{equation}
 is the energy density $\C E$ of \He4 per unit mass.  Hereafter $\< \dots \>$ stands for the  ``proper averaging" which may be time averaging over long stationary dynamical trajectory or/and space averaging in the space-homogeneous case, or an ensemble averaging in the theoretical analysis. Assuming that the Navier-Stokes dynamics are ergodic, all these types of averaging are equivalent.

\begin{figure*}[t]
\begin{tabular}{cc}

    \large (a) Low $T\lesssim T_\lambda/2$, one-fluid model   &  \large  (b) High  $T \gtrsim T_\lambda/2$, two fluid model   \\
 \includegraphics[width=0.51\linewidth]{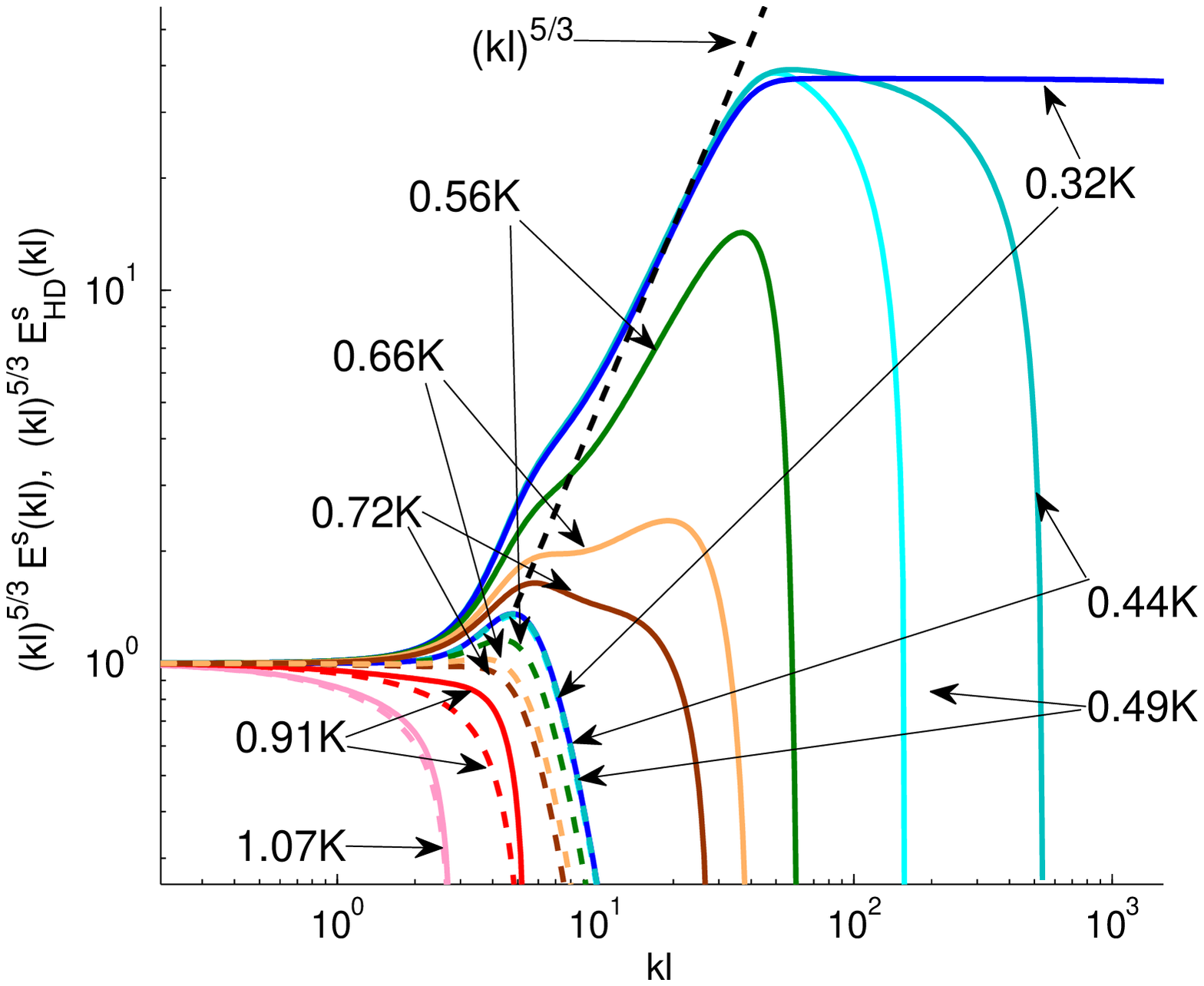}&
 \includegraphics[width=0.49\linewidth]{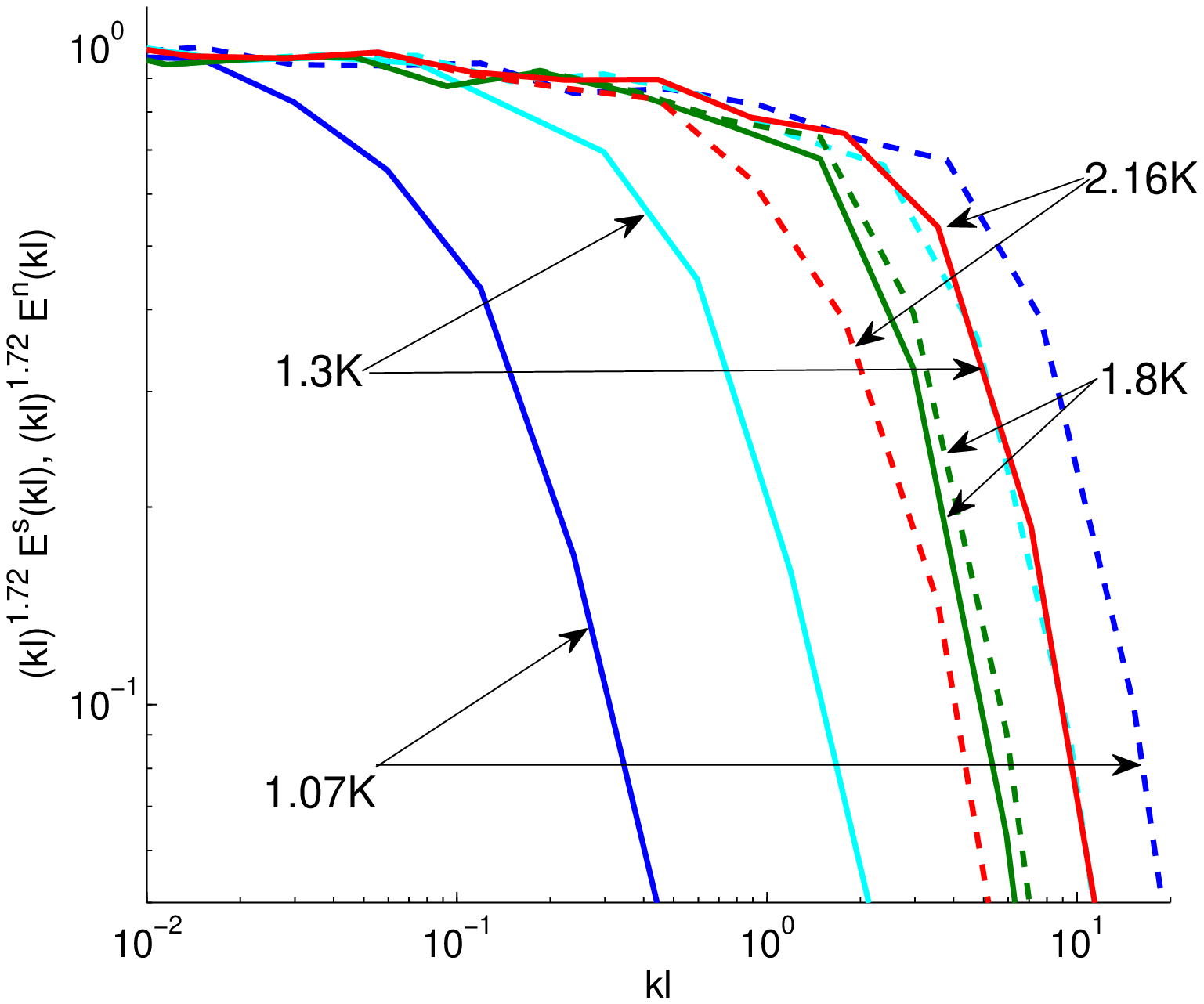} \\

\end{tabular}
\caption{\label{fig1} Color online. Log-Log plots of the  energy spectra in superfluid $^4$He compensated by the inertial range
scaling at different temperatures  (shown as labels). Panel (a): Plots of the (compensated by $k^{5/3}$  and normalized by their
values at $k\ell=0.1$) full superfluid energy spectra (solid lines) $\C E\sp s(\k\ell)$ and their hydrodynamic (large scale) parts
$\C E\sp s \Sb{HD}(\k\ell)$ (dashed lines) for  the one-fluid  model (Sec.\,\ref{s:BB}). Panel (b): Plots of the (compensated by the
anomalous scaling $ k^{1.72}$ and normalized by their inertial range value) shell energies of  the normal fluid component $ |u_m\sp
{n}|^2 = k\C E\sp n (k_m \ell)$ (solid lines)  and of the superfluid component $ |u_m\sp {s}|^2=  k\C E\sp s (k_m \ell)$ (dashed
lines) for the two-fluid shell model (Sec.\,\ref{ss:2fluid}). }
 \end{figure*}

  In the low temperature range, where \He4 consists  mainly of the superfluid component,  we distinguish the spectrum of large scale
  hydrodynamic motions with $k\ell \ll 1$, denoted as $\C E\Sb{HD}\sp s (k)$, from  the spectrum of small scales Kelvin waves (with $k\ell\gg 1$), denoted as $\C E\Sb{KW}\sp s (k)$. The total superfluid energy spectrum is written as
 \begin{subequations}
 \begin{equation}\label{tot-s}
 \C E \sp s (k)\equiv \C E\Sb{HD}\sp s (k)+\C E\Sb{KW}\sp s (k)\ .
 \end{equation}
 In the high temperature range, where the densities of the super-fluid and normal-fluid components are comparable, but Kelvin waves
 are fully damped, we  will distinguish the spectrum of hydrodynamic motions of the superfluid component at large scales as $\C E \sp s (k)$,  from that of the normal-fluid component, $\C E \sp n (k)$, omitting for brevity the subscript ``$\Sb {HD}$".
In this temperature range, the total energy spectrum of superfluid \He4 is written as
 \begin{equation}\label{tot}
 \C E   (k)\equiv \C E  \sp s (k)+\C E  \sp n(k)\ .
 \end{equation} \end{subequations}
  The resulting energy spectra $\C E\sp s\Sb {HD}(k)$,  $\C E\sp s (k)$ and   $\C E\sp n(k)$ for a set of eleven temperatures from
  $T=0.32\,$K to $T=2.16\,$K are shown in \Fig{fig1}.\\

\subsubsection{\label{sss:one-en}Low-temperature one-fluid energy spectra}
First, we discuss the results for the eddy-wave model of superfluid turbulence [cf. Sec.\, \ref{sss:LT-1fluid}], for the low temperature
range $T \lesssim T_\lambda /2 \simeq 1.08\,$K, which is shown in \Fig{fig1}a. These spectra are compensated by a factor $(k\ell)^{5/3}$ such that both the Kolmogorov-Obukhov-41 spectrum $\C E\Sb{HD}\sp s(k)\propto k^{-5/3}$, \Eq{V6-1} (for the hydrodynamic scales $k\ell \ll
1$) and the Lvov-Nazarenko spectrum $\C E\Sb{KW}\sp s (k)\propto k^{-5/3}$, \Eqs{LN-s}   show up as a plateau.
These plateaus are clearly seen for the lowest shown temperature $T=0.32\,$K. Moreover, the full energy
spectrum (solid blue line)    $\C E\sp s(k)$ demonstrates the existence of an important bottleneck
energy accumulation. We observe a large cross over region connecting the HD region $\k\ell< 1$, where
$\C E\sp s(k)\to \C E\sp s\Sb{HD}(k)$ with a much higher plateau  $\C E\Sb{KW}\sp s(k)$ for $k\ell >  50 $  where
$\C E\sp s(k)\to \C E\sp s\Sb{HD}(k)$.

In the cross over region the compensated energy spectrum is close to $(k\ell)^{5/3}$ (cf. the black dashed line), meaning that
$\C E\sp s(k)$ depends on $k$ only weakly.  In this region the energy spectrum is dominated by Kelvin waves, $\C E\sp s(k)\simeq \C E\sp
s\Sb{HD}(k)$, while the energy flux in dominated by the HD eddy motions. Therefore we have here a flux-less regime of Kelvin waves. Without flux the situation resembles
thermodynamic equilibrium,  in which the Kelvin waves energy spectrum corresponds to energy equipartition between the degrees of freedom, i.e. $E\sp s\Sb{KW}(k)\sim$const, as observed.

  For $k\ell > 2 \cdot 10 ^4$ the Kelvin waves energy spectrum at $T=0.32\,$K is suppressed by the mutual friction, as explained in
  Sec.\,\ref{ss:E-HD}. In Fig.\ref{fig1}a this part of the spectrum is not shown; however one sees progressive suppression of the
  energy spectra with temperature increasing from 0.44\,K (around $k\ell \simeq 400$) to $T=1.07\,$K (around $k\ell\simeq 2$).   It
  is important to notice that for $T \geq 0.49\,$K the HD part of the spectrum   is practically temperature independent; only the Kelvin waves energy spectra are suppressed by the temperature, cf. the coinciding dashed lines in \Fig{fig1}a for $T=0.32\,,\ 0.44$ and 0.49\, K.

   For $T>0.5\,$K, the Kelvin wave contributions to the energy spectra are very small -- the solid and the dashed lines for the same temperature are
   fairly close. Finally, the dashed and the  solid lines for $T=1.07\,$K  practically coincide, i.e.  the Kelvin waves  are fully damped. This means that for $T \gtrsim T_\lambda \sim 1\,$K there is no need to account for the Kelvin wave motions on individual vortex lines, and  the full description of the problem is captured by  the coarse-grained HVBK.

 \begin{figure*}[t]
\begin{tabular}{cc}

      \large (a) Low $T\lesssim T_\lambda/2$, one-fluid model   &  \large  (b) High  $T \gtrsim T_\lambda/2$, two fluid model   \\
 \includegraphics[width=0.51\linewidth]{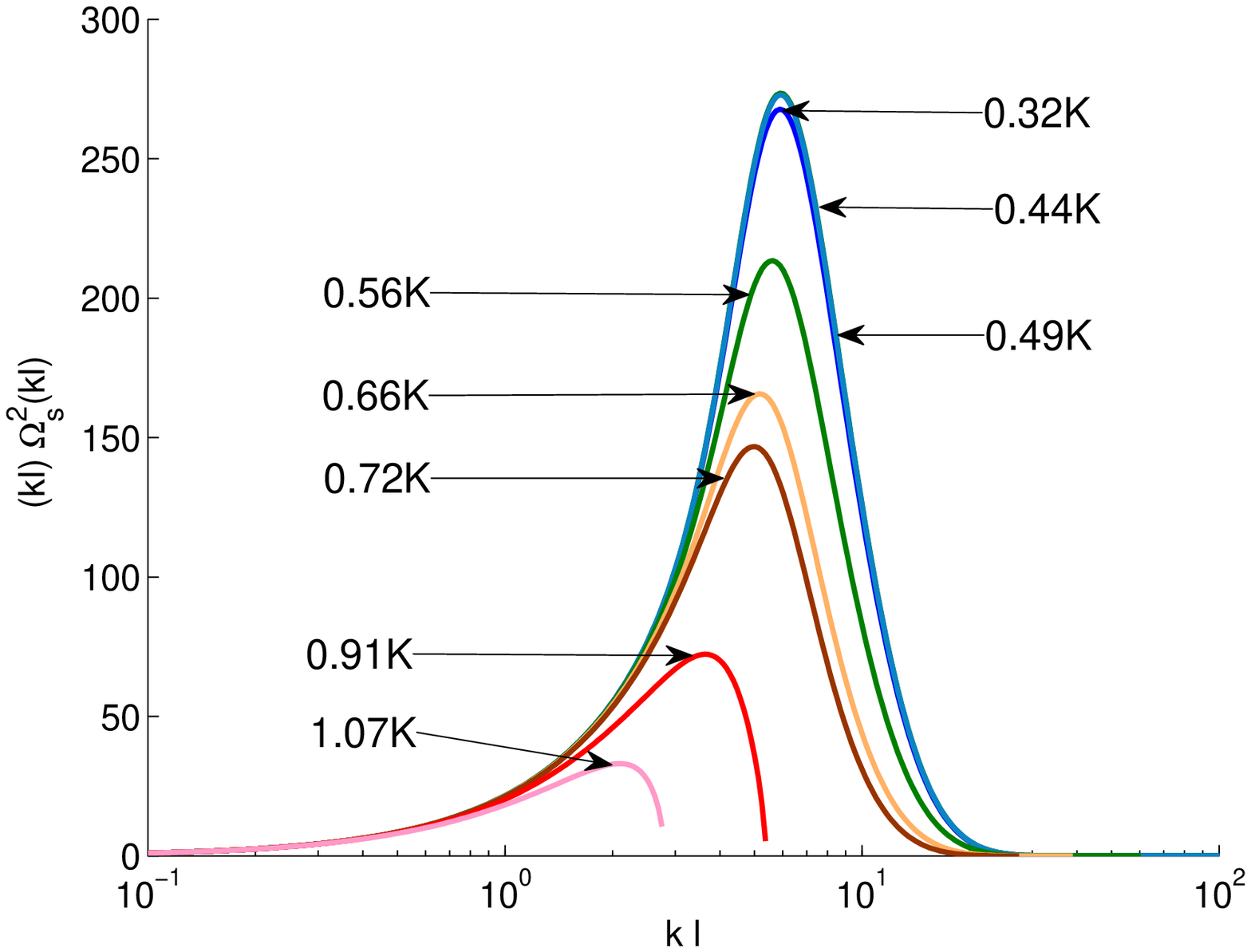}&
 \includegraphics[width=0.49\linewidth]{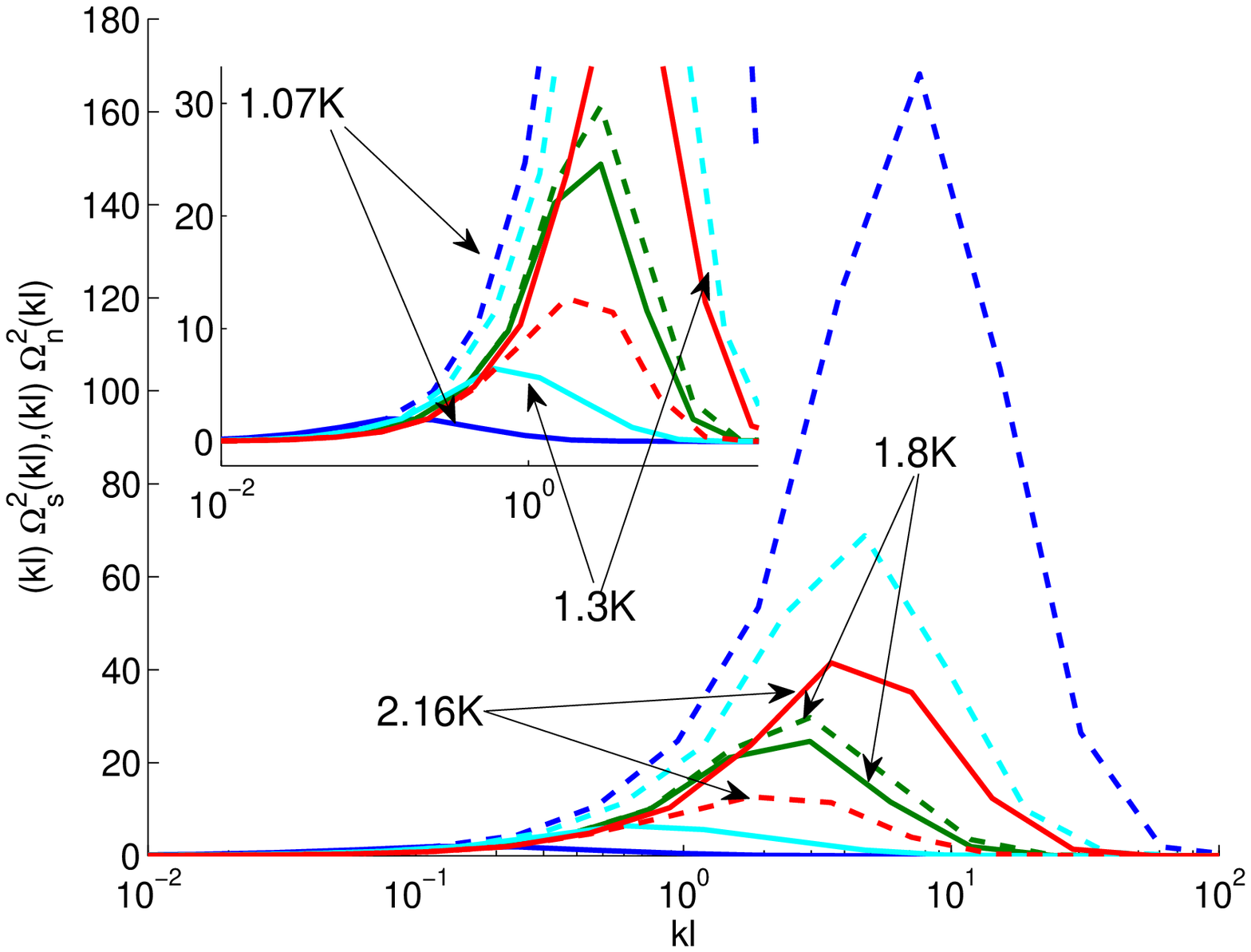} \\

\end{tabular}
\caption{\label{fig2} Color online. Color online. Linear-log plots of the  vorticity spectral  densities, $(k\ell)\,
\Omega^2(k\ell)$, normalized by their $k\ell=0.1$ values,  at different temperatures. The total  mean-square vorticity $\< \omega^2
\>$   is proportional to the area under the plot.  Panel (a): Superfluid vorticity  spectra  $(k\ell)\, \Omega \sb s ^2(k\ell)$  at
low temperatures in the one-fluid model,  Sec.\,\ref{s:BB}.
Panel (b):   Normal and superfluid vorticity  spectra  $(k\ell)\,  \Omega \sb n^2(k\ell)$ (solid lines) and $(k\ell)\,  \Omega \sb
s^2(k\ell)$ (dashed lines)  at high temperatures in the two-fluid model,  Sec.\,\ref{ss:2fluid}. }
 \end{figure*}

\subsubsection{\label{sss:two-en}High-temperature two-fluid energy spectra}

  The energy spectra, obtained with the  Sabra-shell  model form  of HVBK  equations\,\eqref{smA} and \eqref{Sabra}, for temperatures $T \gtrsim T_\lambda /2 \simeq 1.08\,$K,  are shown \Fig{fig1}b  for $T=1.07\,$K (in blue), $T=1.3\,$K (in magenta), $1.8\,$K
 (in green) and $T=2.16\,$K (in red).   The lowest temperature in this two-fluid approach,  $T=1.07\,$K,  was chosen for  comparison
 with the highest temperature  $1.07\,$K in the one-fluid approach; see \Fig{fig1}a. At $T=1.3\,$K, which is a frequently used temperature in
 numerical simulations of superfluid turbulence, the normal fluid component is not negligible ($\rho\sb n /\rho \simeq 0.045$), and the normal fluid kinematic
 viscosity is still much larger than that of the superfluid: $\nu\sb n/ \nu'\sb s \simeq 23$. For $T=1.8$, when $\rho\sb n /\rho  \simeq
 0.3$, the kinematic viscosities are close to each other (see Tab.\,\ref{t:1}).  At higher temperatures the normal fluid components
 play more and more important role until they dominate at $T>2.0\,$K, when  $\rho\sb n > \rho\sb s$. At the highest temperature in this
 simulation,
 $T=2.16\,K$, close to $T_\lambda$, we have $\rho\sb n \approx 0.9 \rho$, and the effective superfluid kinematic viscosity $ \nu'\sb
 s$ is even larger than $\nu\sb n$.

 Shell-model simulations reproduce intermittency effects and therefore the scaling exponent $\xi_2$ of the energy spectra $\C
 E(k)\propto k^{-\xi_2}$  slightly  differs from the KO-41 prediction, $\xi_2\ne 5/3$. For the chosen   shell-model
 parameters\cite{BLPP-2013,Sabra}  $\xi_2\approx  1.72$ which is quite close to the experimental observations. For better comparison with the low-temperature  one-fluid results of \Fig{fig1}a,  we show  in \Fig{fig1}b the normal (solid lines) and superfluid (dashed lines) energy spectra $\C E\sp n(k\ell)$ and $\C E\sp s(k\ell)$, compensated by $(k\ell)^{1.72}$ so that they exhibit a plateau in the inertial interval of scales.

As expected, for $T=1.8\,$K, when   $\nu\sb n\approx \nu\sb s'$, the superfluid and normal fluid spectra are very close, and similar
to the spectra of classical fluids. In the inertial range they demonstrate the  anomalous behavior $\C E\sp s \propto \C E\sp n \propto k^{-\xi_2}$  $  [|u_m\sp
{n}|^2\propto k_m^{1-\xi_2}$] with the scaling exponent $\xi_2 \approx 1.72$.  Moreover, due to  the strong coupling between the normal and superfluid component (discussed   below  in \Sec{ss:intro-D}) the energy fluxes in both components
are equal (see, e.g. \Fig{fig4}), and therefore  the energies are equal  in the inertial interval as well, $\C E\sp s(k)=\C E\sp
n(k)$. Non-trivial behavior occurs only in the inertial-viscous crossover region; therefore the inertial interval is not shown in
\Fig{fig1}.

For $T=1.07$, when $\nu\sb n \simeq 180 \,\nu'\sb s$, the viscous cutoff of the normal fluid's spectrum, $k\sb {max,n}$, occurs at
much smaller $k$ than the cutoff of the superfluid spectrum,  $k\sb {max,s}$. To estimate the ratio $k\sb {max,s}/k\sb {max,n}$,
notice that in the KO-41 picture of turbulence $k\sb {max}$ may be found by balancing the eddy-turnover frequency,
\begin{equation}\label{gamma} \gamma(k)\simeq \varepsilon ^{1/3}k^{2/3}\simeq k^{3/2}\sqrt {\C E(k)}\,,
 \end{equation}
 with the viscous dissipation frequency $\nu k^2$. This gives the well known result \begin{equation}\label{kmax}
 k\sb{max}\simeq \ve^{1/4}/\nu^{3/4}\ .
  \end{equation}
  In our case $\ve\sb s=\ve \sb n$. Therefore, neglecting the energy exchange between the super- and the normal-fluid components, we
  get an estimate:
\begin{equation}\label{estim1}
\frac{k\sb {max,s}}{k\sb {max,n}}\simeq \Big( \frac{\nu\sb n}{\nu'\sb s}\Big)^{3/4}\ .
\end{equation}
 For $T=1.07$, when $\nu\sb n \simeq 180 \,\nu'\sb s$ this gives  $k\sb {max,s}/k\sb {max,n}\simeq 50$---in a good agreement with the
 result in \Fig{fig1}b.
For $T=1.3\,$K, the ratio of the viscosities is smaller (about $\simeq 23$, see Tab.\,\ref{t:1}). Therefore the  difference in
cutoffs is   less pronounced. As expected,  for $T=2.16\,$K,  when $\nu\sb n \simeq 0.2 \nu'\sb s$ the situation is the opposite, and
the superfluid component is damped at a smaller $k$ than the normal one.

Notice that there is no bottleneck energy accumulation in the spectra  (see \Figs{fig1}b) obtained using the shell model
approximation of the
\emph{gradually damped} HVBK equations.
This  is qualitatively different from the results  of the \emph{truncated} HVBK model\,\cite{22}, which demonstrated a very pronounced
bottleneck both in the normal and the superfluid components, e.g.  at   $T=1.15\,$K.  The latter would lead to a huge contribution to
the mean square superfluid vorticity $\< |\omega_s|^2\>$ and, as a result,  to a very small effective Vinen's viscosity $\nu'\sb s$.
This would definitely contradict the experimental observation shown in \Fig{fig5}. We will discuss this issue in greater detail in
\Sec{ss:intro-E}.
 \subsection{\label{ss:intro-C}Temperature dependence of the vorticity  spectra   in turbulent $^4$He}

 At this point we cannot compare our predictions for energy spectra with experimental observations,
 especially in the cross-over and in the small scale regions. This stems from the lack of small probes, see
cf. the review~\cite{BLR}. On the other hand, the attenuation of second sound or ion scattering may be used to measure the  mean
vortex line density $\C L= 1/\ell^2$ in \He4 or even its time and space dependence~\cite{BLR}. In  turn, the value $\C L^2$
can be expressed in terms of the mean-square superfluid vorticity $\< |\omega\sb s|^2\>$  via the quantum of circulation
$\kappa$~\cite{Vin-2001}:
\begin{equation}\label{Vin}
\< |\omega\sb s|^2\>\approx (\kappa \C L)^2\ .
 \end{equation}
Therefore, the information about the vorticity  is very important from the viewpoint of comparison with available and future experiments.

By analogy with the energy spectra\,\eqref{def0}, let us define the power spectra of vorticity  $\Omega^2(k)$  so that the
mean-square vorticity  $\< |\omega|^2 \>$ is given by the integral:
\begin{equation}\label{Om}
 \< |\omega|^2 \>= \int\limits_0^\infty \Omega^2(k) \, dk= \int\limits_0^\infty  k\  \Omega^2(k)  \, d\ln k\ .
\end{equation}
In isotropic incompressible turbulence $\Omega^2(k)= 2 k^2 \C E\Sb {HD}(k)$.   Therefore we define
  \begin{eqnarray}\label{def7}
  \Omega^2\sb s (k)= 2 k^2 \C E \sp s\Sb{HD}(k)\,,  \quad \Omega^2\sb n (k) = 2 k^2 \C E \sp n(k)\,.
\end{eqnarray}
For brevity, we omit the subscript ``$\Sb {HD}$" for the normal component;  $\C E \sp n\Sb {HD}(k) \Rightarrow \C E \sp n (k)$.

Plots of $k\,\Omega^2\sb{s,n}(k)$ for different temperatures  are shown in Figure\,\ref{fig2}.  According to \Eq{Om}, the area under
these plots is proportional to the total mean square vorticity, $\< |\omega\sb {s,n}|^2\>$. Fig.\,\ref{fig2}a  shows the  results for the eddy-wave model  (the corresponding energy spectra  for the same temperatures are shown in \Fig{fig1}a). One sees
that  the largest (and temperature independent) value of  $\< |\omega\sb {s}|^2\>$  is reached for
$T < 0.49\,$K: plots for $T=0.32\, \ 0.44\, $ and $0.49\,$K practically coincide.   Accordingly, the temperature range $T < 0.49\,$K
may be considered as zero-temperature limit with the maximal value of  $\< |\omega\sb {s}|^2\>$ (and correspondingly, the smallest
value of $\nu\sb s(T)$, as we will discuss later).  At temperatures above 0.5K the area under the plots decreases (and
correspondingly, $\nu\sb s(T)$ increases).

In Fig.\,\ref{fig2}b we show
 vorticity spectra $(k\ell)\<|\omega\sb {s,n}|^2(k\ell)\>$ of  the normal-fluid (solid lines) and the superfluid components (dashed
 lines) for different temperatures   obtained in the framework of  the Sabra-shell model  (the corresponding spectra are shown in Fig.\,\ref{fig1}b).  Again, the area under the plots is proportional to the
 total mean square vorticity $\<|\omega\sb{n,s}|^2\>$. One sees that for the lowest temperature $T=1.07\,$K the normal fluid
 vorticity (blue solid line) is fully suppressed by the huge normal viscosity, while the superfluid vorticity is very large. At this
 temperature one can describe the superfluid $^4$He in the range
 of scales  $k\ell \sim 1$ using a one-fluid approximation with zero normal-fluid velocity.  This provides the main contribution to the vorticity. To some extent,  this situation persists up to
 $T\approx 1.3\,$K, when the superfluid vorticity is still larger than the normal one, see  Fig.\,\ref{fig1}b. As expected, for
 $T\simeq 1.8$, when the normal and superfluid viscosities are compatible, the normal and superfluid vorticities are very close. For
 these and higher  temperatures the analysis of our problems definitely calls for a two-fluid description.

 \subsection{\label{ss:intro-D} Correlations of normal and superfluid motions and energy exchange between components}
 \begin{figure}[t]
 \begin{tabular}{c}
   \large  High  $T \gtrsim T_\lambda/2$, two-fluid model \\
  \includegraphics[width=0.95\linewidth]{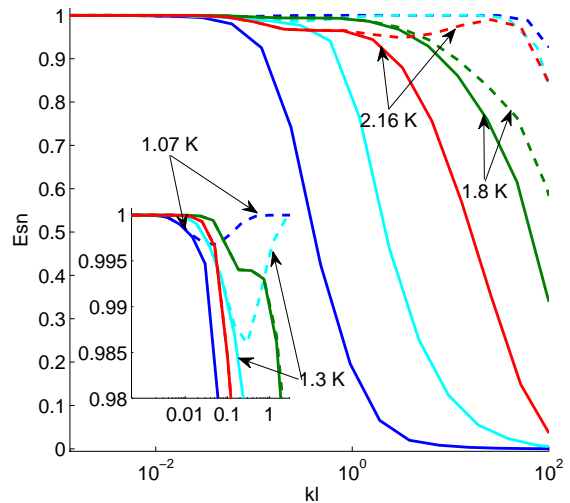} \\
  \end{tabular}
  \caption{\label{fig3} Color online.
    Cross-correlation coefficients $K_1(k\ell)$,  \Eq{cross1} (solid lines) and  $K_2(k\ell)$ , \Eq{cross2} (dashed lines) for
    different temperatures. Color code is the same as in \Fig{fig1}b and \Fig{fig2}b:  $T=1.07\,$K -- blue, $T=1.3\,$K -- cyan,
    $T=1.8\,$K -- green,  and $T=2.16\,$K -- red. }
\end{figure}
\subsubsection{Correlations of the normal and superfluid velocities}
 It is often  assumed (see e.g. Ref.\cite{VinenNiemela})  that the normal and superfluid velocities are ``locked" in the sense that
 \begin{equation}\label{lock}
 \B u\sb n (\B r,t)= \B u\sb s (\B r,t)\,,
 \end{equation}
 (at least in the inertial interval of scales).
 For quantitative understanding to which extent this assumption is  statistically valid we consider the simplest possible case of
 stationary, isotropic and homogeneous turbulence. Here we introduce a
 cross-correlation function (in 1D $k$-representation) of the normal and the superfluid velocities   $\C E\sb{ns}(k)$.   This
 correlation function is defined using the simultaneous, one-point cross-velocity correlation  $\< \B u \sb n (\B r,t)\* \B u \sb s
 (\B r,t)\>$ similarly to \Eq{def0}:
\begin{equation}\label{def1c}
\int \C  E\sb{ns}(k) \, dk= \frac 12\< \B u \sb n (\B r,t)\* \B u \sb s (\B r,t)\> \ .
\end{equation}
 If, for example,  motions of the normal and the superfluid components at a given $k$ are completely correlated, then $\C E\sb {ns}
 (k)= \C E\sb n (k)= \C E\sb s (k)$. If this is true for all scales, then \Eq{lock} is valid.

 It is natural to normalize $\C E\sb{ns}$ by the normal and the superfluid energy densities, $\C E\sb{n}$ and $\C E\sb{s}$. This can
 be reasonably done in one of two  ways:
 \begin{subequations}\begin{eqnarray}\label{Def1}
 \C K_1(k)\=\frac{2\, \C  E\sb{ns}(k)  } {\C  E\sb{n}(k)+\C  E\sb{ s}(k)}\,, \\
 \hbox{or} \;\;\; \C K_2(k)\=\frac{ \C  E\sb{ns}(k)  } {\sqrt{\C  E\sb{n}(k)\cdot \C  E\sb{ s}(k)}}\ .
 \end{eqnarray}\end{subequations}
 Both coefficients are equal to unity for fully locked superfluid and normal   velocities, \Eq{lock},   and both vanish if the
 velocities are statistically independent. However, if $ \B u\sb  n (\B r,t)=C\B u\sb s (\B r,t)$, with $C \ne  1$
then $\C K_1 (k)= 2C / (C^2+1)<1$, but still $\C K_2 (k)= 1$. In any case $\C K_1 (k) \leqslant \C K_2 (k)$.

In shell models, the coefficients $\C K_1(k)$ and $\C K_2(k)$ can be written as follows:
\begin{subequations}\label{cross}
\begin{eqnarray}\label{cross1}
K_1 (k_m\ell) &\equiv &   \frac{2\,\mbox{Re}\<   v_m\sp {s*}  v_m\sp n\>}
 { \< v_m\sp {s*} v_m\sp s\> + \< v_m\sp {n*}  v_m\sp n\> }\,, \\ \label{cross2}
 K_2 (k_m\ell)&\equiv&   \frac{\mbox{Re}\<  v_m\sp {s*}  v_m\sp n\>}
 {\sqrt{\< (v_m\sp {s*} v_m\sp s\>  \< v_m\sp {n*}  v_m\sp n\>}}\ .
\end{eqnarray}
\end{subequations}
These objects are shown in \Fig{fig3}.
At first glance, it is surprising that the correlations   $K_2 (k_m\ell)$ (dashed lines in \Fig{fig3}) for $T \lessgtr 1.8\,$K
persist for much larger wave vectors than $K_1 (k_m\ell)$, approaching $k_m\ell \sim 10^2$. For example, for $T=1.07\,$K (blue lines)
$K_1(k_m\ell)$ vanishes at $k_m\ell \simeq 1$, while  $K_2(k_m\ell)> 0.95$ all the way up to
$k_m\ell \simeq 100$. In this range of scales ($1 \lesssim k_m\ell \lesssim 100$)  $v_m \sp n \ll v_m \sp s$, but  $v_m \sp n
(t)\propto  v_m \sp s(t)$, meaning that strongly damped normal velocity does not have its own dynamics and should be considered as
``slaved" by the superfluid velocity.  The damped velocity (normal or superfluid)  at any temperature   $T \lessgtr 1.8\,$K would
follow this ''slaved" dynamics.

  A model expression  of  the cross-correlation $\C E\sb {sn}$ in terms of the self-correlation functions  $\C E\sb {s}$ and
  $\C E\sb {n}$ was found in \Ref{L199}. In current notations it reads:
\begin{eqnarray}\label{lim1}    \C E\sb{sn}(k) = \frac {\a\,  \bar\o \sb s [\r \sb n \,  \C E\sb n(k)+ \r \sb s \,  \C E\sb s(k) ] }
{\a \,\bar\o \sb s \r+ \r\sb n [(\nu'\sb s+ \nu\sb n)\, k^2 + \g\sb n(k)+ \g \sb s(k)]   }\,, \end{eqnarray}
where the   characteristic interaction frequencies (or turnover frequencies) of eddies in the normal and superfluid components, $\g
\sb n(k) $ and $  \g \sb s(k) $,  are given by \Eq{gamma} and $\bar\o \sb s $  is defined as:
\begin{equation}\label{baros} \bar\o \sb s\= \sqrt {\<|\omega\sb s|^2 \>}\ .
  \end{equation}
  The derivation of \Eq{lim1} in \Ref{L199} involves diagrammatic perturbation approach and is rather cumbersome. However the
  simplicity of the final result~\eqref{lim1} motivated us to re-derive it in a simple and transparent way which is presented in the  Appendix.

Let us analyze first \Eq{lim1} in the inertial interval of scales, where according to \Fig{fig1}b, $\C E \sb s=\C E \sb n$ and the terms
with the viscosities in the denominator may be neglected. In this case
\begin{eqnarray}\label{case1}
\C K _1(k)&\to&   \Big [ 1 + \frac{2 \rho\sb n\gamma\sb s(k)}{\alpha \, \r \,\bar \omega \sb s} )\Big ]^{-1}\\ \nn &\simeq&  \Big [ 1
+ \frac{2\, \rho\sb n }{\alpha \, \r } \Big ( \frac{k}{k\sb{max,s}}\Big )^{2/3}\Big ]^{-1}\,,
\end{eqnarray}
where the viscous cutoff of the superfluid inertial interval $k\sb{max,s}$ is given by  estimate\,\eqref{kmax}.
First of all we see that the correlation coefficient is governed by the dimensionless parameter $\alpha \r/\r\sb n$ which involves
the mutual friction coefficient $\alpha$, as expected. What is less expected,  is that this parameter, according to Tab.\,\ref{t:1},
depends on the temperature only weakly and is close to unity. Therefore, in the inertial interval $k\ll k\sb{max,s}$ we have:
\begin{eqnarray}\label{case2}
\C K _1(k)&\simeq&    \Big [ 1 + \frac{2\, \rho\sb n }{\alpha \, \r } \Big ( \frac{k}{k\sb{max,s}}\Big )^{2/3}\Big ]^{-1}\\ \nn
&\simeq &  1 - \Big ( \frac{k}{k\sb{max,s}}\Big )^{2/3} \,,
\end{eqnarray}
and this expression is very close to unity. In the other words, in the inertial interval we expect the full locking of the normal and
the superfluid velocities for all temperatures. This prediction fully agrees with the observations in \Fig{fig3}.

Consider now case $T=1.07\,$K, when $k\sb{max,s}\simeq 50\, k\sb{max,n}$ according to the data in \Fig{fig1}b and
estimate\,\eqref{estim1}. For $ k\sb{max,n}<k < k\sb{max,s}$ we have: \begin{equation}\label{estim2}
\  \C E \sb s(k) \gg \C E \sb n(k)\,, \ \ \mbox{and} \ \ \nu \sb n \gg \g \sb s(k) \gg \g
\sb n(k)\gg \nu'\sb s \ .
\end{equation}
Then \Eq{lim1} simplifies  to the following form,
\begin{equation}\nn
\C K _1(k)\simeq    \Big [ 1 +   \frac{  \rho\sb n }{\alpha \, \r }  \frac{\nu \sb n k^2}{\bar\o \sb s}\Big ]^{-1},
\end{equation}
and it may be analyzed as follows:
\begin{equation}\nn
\C K _1(k)\simeq \Big [ 1 +   \frac{  \rho\sb n }{\alpha \, \r }  \frac{\nu \sb n k^2}{\bar\o \sb s}\Big ]^{-1}\simeq    \Big [ 1 + \frac{\nu \sb n k^2}{\nu'  \sb s\,  k^2 \sb{max,s}}   \Big ]^{-1}
\end{equation}
 { Using \eq{estim1}, for $\ k\sim k\sb{max,s}$ we get
 \begin{equation}\label{case3}
 \C K _1(k) \simeq \Big [ 1 +   \frac{k^{4/3} \sb{max,s}}{    k^{4/3} \sb{max,n}}   \Big ]^{-1}  \sim   \frac{    k^{4/3} \sb{max,n}} { k^{4/3} \sb{max,s}}\ll  1\,,
\end{equation}

 We see that the velocities decorrelate in the interval $ k\sb{max,n}<k < k\sb{max,s}$,  as expected.

Estimating $\C K _2(k)$ in the regime\,\eq{estim2} is less simple, because it requires  knowledge of the ratio $\C E\sb s (k)/ \C E\sb n
(k)$ in  terms of $\nu\sb n, \nu'\sb s, \gamma_n(k)$ and $\gamma_s(k)$. Instead, we can directly use \Eq{NSE3b}, which in  regime
\,\eq{estim2} may be simplified (in the $(\B k,t)$-representation) as follows:
 \begin{equation}\label{rat1}
  u \sb n (\B k, t)= \frac{\a\, \r \sb s\,\bar \o \sb s}{\r \sb n \nu\sb n k^2} \,   u \sb s (\B k, t)\ll   u \sb s (\B k, t)\ .
\end{equation}
i.e. $ u \sb n (\B k, t)$ is slaved by $ u \sb s (\B k, t)$.} Equation\,\eq{rat1}
immediately gives  $\C K _1(k)\ll 1$, but  $\C K _2(k)= 1$ in full agreement with our results in \Fig{fig3}. In particular, this
means that our simple model of correlations between  $\B u \sb n$ and $\B u \sb s$, suggested in the Appendix, quantitatively correctly
reflects the basic physics of this phenomenon.
\begin{figure}[t]
 \begin{tabular}{c}
   \large High  $T \gtrsim T_\lambda/2$, two fluid model \\
  \includegraphics[width=0.95\linewidth]{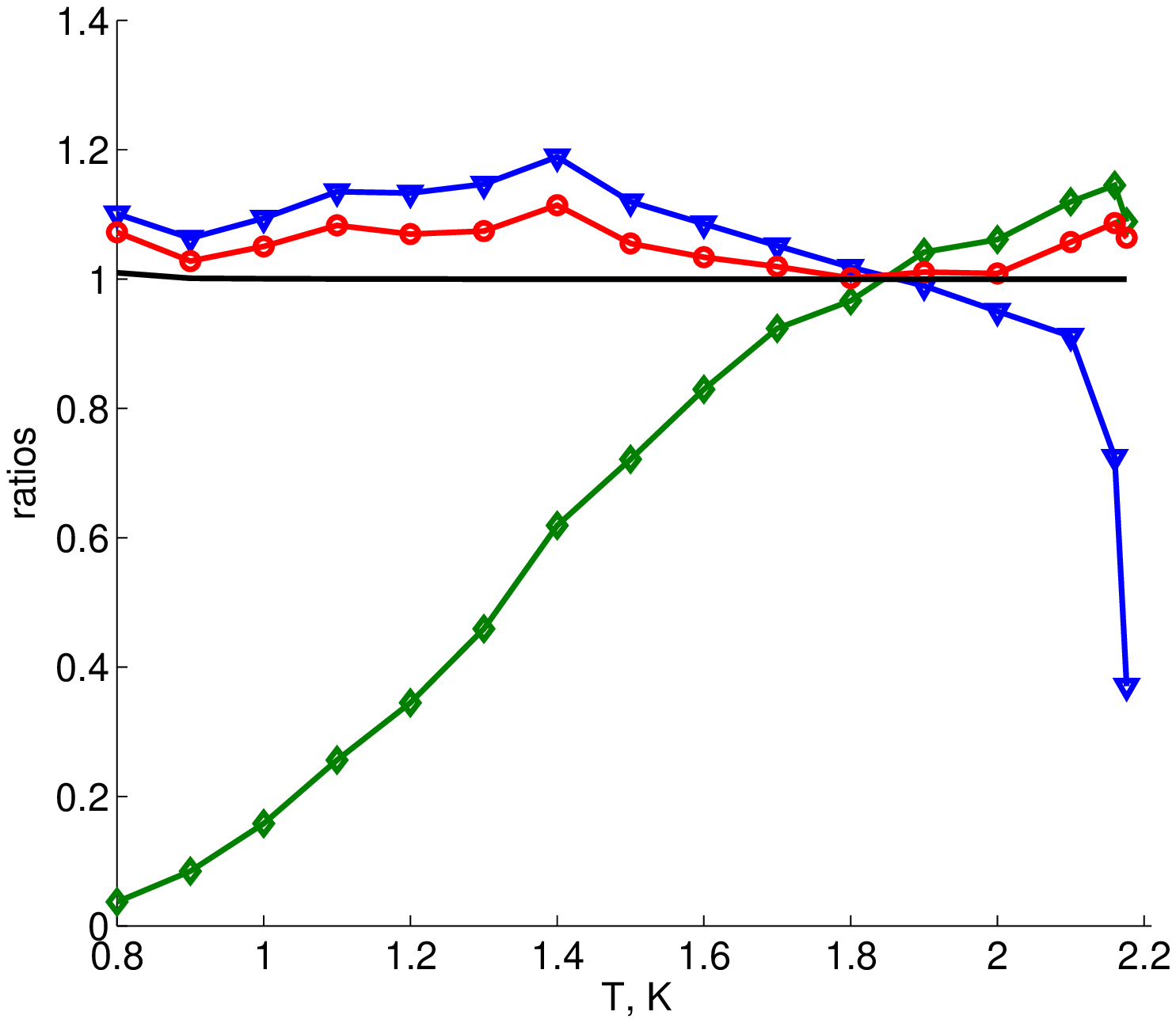} \\
  \end{tabular}
  \caption{\label{fig4} Color online. Temperature dependence of the ratios: $ \ve  \sb s/ \ve \sb n$ -- horizontal black line;  $R\sb
  s$,  \Eq{ratsA}, -- blue line with triangles,   $R\sb n$,  \Eq{ratsA}, -- green line with diamonds, $R\sb {s+n}$, \Eq{ratsB}, --
  red line with circles. }
\end{figure}

\subsubsection{Energy dissipation and  exchange due to   mutual friction}

Strong coupling of the normal  and the superfluid velocities suppresses the energy dissipation and the energy exchange between the
normal   and the superfluid components  caused by the mutual friction (which is proportional to $\B u\sb s - \B u\sb n$, \Eq{NSEf-a}).
Nevertheless, some dissipation due to the mutual friction is still there. Consider the  ratio of the total injected energy to the
total energy dissipated due to the viscosity in the normal and the superfluid components:
\begin{subequations}\label{rats}
 \begin{equation}\label{ratsA}
 R\sb {s+n}= \frac{\rho \sb s \ve \sb s+\rho \sb n \ve \sb n}{\rho \sb s\nu'\sb s \< |\omega\sb s|^2\>+\rho \sb n\nu \sb n \<
 |\omega\sb n|^2\>}\ ,
 \end{equation}
a quantity plotted in \Fig{fig4} (red line with circles). Here $\ve \sb n$ and $\ve \sb s$ are the inertial range normal and
superfluid energy fluxes.
This ratio exceeds unity by about 10\%, meaning that $\sim 10\%$ of the injected energy is dissipated by the mutual friction. As
expected, this effect disappears at $T\approx 1.8\,$K, when the effective superfluid and normal fluid kinematic viscosities are
matching (and therefore $\B u\sb s \approx \B u\sb n$).

The mutual friction has a significantly more important influence on the energy exchange between the normal and the superfluid
 components. The energy exchange can be quantified by a similar ratio defined for each fluid component,
 \begin{equation}
 R\sb s=\frac{\ve \sb s}{\nu'\sb s \< |\omega\sb s|^2\>}\,, \quad  R\sb n= \frac{\ve \sb n}{\nu \sb n \< |\omega\sb n|^2\>}\,,
 \label{ratsB}
 \end{equation}
 \end{subequations}
shown  by a green line with diamonds and a blue line with  triangles respectively in \Fig{fig4}.
At the lowest shown temperature $T=0.8\,$K, we have $R\sb n < 0.1$ meaning that only  about 10\% of the energy density (per unit
mass) which is dissipated by the normal fluid component comes from the direct energy input. The rest $\simeq 90\%$ of the energy density dissipated by
viscosity (at large $k$) was transferred from the superfluid component by the mutual friction. This is because for $T
\lesssim 1.8\,$K, we have $\nu\sb n > \nu'\sb s$ and therefore the normal velocity becomes more damped at lower wavenumbers than the
superfluid velocity,
  see \Fig{fig1}b. Such an energy transfer by mutual friction from the superfluid component to the normal one increases  $R\sb s $ (blue
  line with triangles) above  unity. This effect is smaller than the one for $R\sb n$ because at low temperatures $\rho\sb s \gg \rho
  \sb n$ and the energy per unity volume $\rho \sb s \ve \sb s+\rho \sb n \ve \sb n$ is approximately conserved. As expected, there
  is no energy exchange between the components at   $T \approx 1.8\,$K, when $\nu\sb n = \nu'\sb s$ and $\B u\sb s = \B u\sb n$). At
  this temperature $R\sb s= R \sb n = R \sb {n+s}=1$.  Again, as expected for $T > 1.8\,$K, when $\nu\sb n < \nu'\sb s$ (see
  Tab.\,\ref{t:1}) we have $R\sb s > 1$,  $R \sb n < 1$ meaning that the energy goes from the less damped normal component to the
  more damped superfluid one.

To understand  why the energy exchange due to the mutual friction is larger than the energy dissipation by the mutual friction,
notice that  the  energy exchange is proportional to  the (small) velocity difference, $  \<\B u \sb {n} \cdot ( \B u\sb n - \B u_s ) \> $, while the energy dissipation is proportional to the square of this parameter, $  \< |\B u \sb n - \B u\sb s |^2\>
$.

\begin{figure*}[t]
\begin{tabular}{cc}
 \large  Low $T\lesssim T_\lambda/2$, one-fluid model~~    & \large   ~~ High  $T \gtrsim T_\lambda/2$, two-fluid model \\
\end{tabular}
 \includegraphics[width=0.85\linewidth]{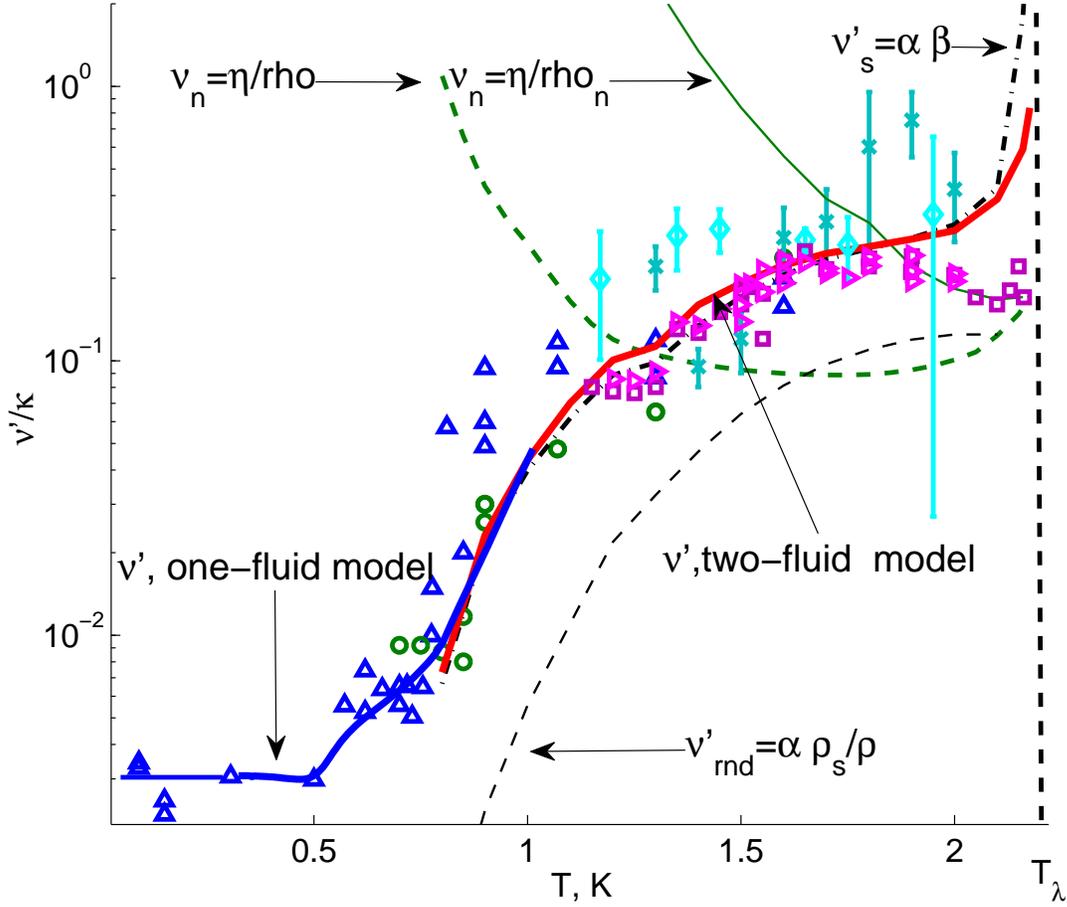}
\caption{\label{fig5} Color online.  Comparison of the experimental,  numerical and analytical results for  the temperature
dependence of the effective kinematic viscosities: {Blue triangles -- Manchester spin-down experiments~\cite{Manchester-exp}; Green empty circles -- Manchester ion-jet experiments\,\cite{Golov2}; sea-green diamonds with error-bars  -- Prague counterflow experiments~\cite{Ladik}; cian crosses with error-bars -- Prague decay  in grid co-flow experiments\,\cite{Ladik1}; Magenta  empty squares -- Oregon towed grid experiments~\cite{stalp};Pink right triangles--Oregon towered grid experiments~\cite{nimiela2005}}.
Solid green line --  experimental results\,\cite{DonnelyBarenghi98} for the normal-fluid kinematic viscosity $\nu\sb
n=\mu/\rho\sb{n}$ (normalized by the normal-fluid density); Dashed green line -- He-II kinematic viscosity $\nu\equiv \mu/\rho$,
(normalized by the total density) -- see also Tab.\,\ref{t:1}. {Thin black dash line -- effective viscosity for the random vortex
tangle $\nu'\sb{rnd}$, estimated by \Eq{nurnd}}; Thick dot-dashed black  line --  the Vinen-Niemela estimate\,\cite{VinenNiemela} of
the effective superfluid  viscosity, $\nu'\sb s$, given by \Eq{nus}.  Blue solid line -- $\nu'(T)$ at     $T<1.1\,$K from  numerical
solution of Eqs.\,\eqref{A2} for the one-fluid differential model of gradual eddy-wave crossover;
Red solid line -- $\nu'(T)$ at    $T>0.9\,$K from numerical simulations in Sec.\,\ref{ss:2fluid}  of gradually damped two-fluid HVBK
\Eqs{NSE} in the  Sabra shell-model approximation\,\eqref{SM}.    }
 \end{figure*}


 \subsection{\label{ss:intro-E}Temperature dependence of the effective superfluid viscosity  in $^4$He}
  The temperature-dependent effective (Vinen's) viscosity $\nu'(T)$  is defined\,\cite{stalp} by the relation between the rate of
  energy-density (per unit mass) flux into turbulent superfluid,  $\varepsilon$,  and the vortex-line density, $\C L$:
\begin{equation}\label{oT}
\varepsilon =  \nu' (T)(\kappa \C L)^2 \approx \nu' (T) \< |\omega\sb s|^2\>\ .
\end{equation}
According to \Eq{Om}, $\< |\omega\sb s|^2\>$ is proportional to the area under the plots $k\,\Omega^2\sb s (k) $ vs. $\log k$, shown
in Fig.\,\ref{fig2}  and discussed in \Sec{ss:intro-C}.  These  results allow us to determine  the  viscosity
$\nu'(T)$ (analytically and numerically) in the entire temperature range from $T\to 0$ up to $T\to T_\lambda$.

\subsubsection{\label{sss:T-to-0} Low temperature range $T \lesssim T_\lambda /2 $}

 Consider first  the temperature dependence of $\nu'(T)$ in  the low temperature range $T \lesssim T_\lambda /2 \approx 1.1\,$K,
 shown in  \Fig{fig5} by the solid blue line.  This dependence is  found in  \Sec{s:BB} in the framework of one-fluid model of
 gradual eddy-wave crossover \Eqs{dif-m}.   As we mentioned,  the largest (and temperature independent) value
 of $\< |\omega\sb s|^2\>$  (at fixed value of $\ve\sb s$)  is reached for $T<0.49\,$K. Accordingly, the temperature range
 $T<0.49\,$K may be considered as a  zero-temperature limit, at which $\nu'(T)$ reaches its smallest value.  The results of the
 Manchester spin-down experiment\,\cite{Manchester-exp}  are  temperature independent as well (within the natural
 scatter of the data). The particular value of $\nu'\sb{exp}(T\to 0)\simeq 0.003 \, \kappa$ found in these experiments is probably
 accurate up to a numerical factor ${(\frac 13 \div 3)}$ due to uncertainty   in the  determination of the  outer scale of
 turbulence, taken in  Ref. \cite{Manchester-exp} for simplicity as the size of the cube. Our low-temperature,  one-fluid model\,\eqref{dif-m}  involves one
 fitting parameter, which determines the crossover scale in the blending function\,Eq.\eqref{est5}. This parameter affects the
 resulting value of $\nu'\sb{mod}(T\to 0)$ and was chosen such as to meet its accepted experimental value $\nu'\sb{exp}(T\to 0)\simeq
 0.003 \, \kappa$.

 At   temperatures above 0.5\,K, the area under the $k \Omega^2\sb s(k)$ plots in \Fig{fig2}a  become smaller and smaller. This is
 caused by the  suppression of the Kelvin wave spectra, which is more pronounced at larger  temperature, as seen in \Fig{fig1}a.  The value of $\< |\omega\sb s|^2\>$ decreases with the temperature resulting in a progressive increase  in $\nu'(T)$ as shown by the solid blue line in
 \Fig{fig5} together with the experimental (Manchester spin-down\cite{Manchester-exp} and ion-jet\cite{Golov2}) values of
 $\nu'\sb{exp}(T)$.  There is a reasonably good agreement between the temperature dependence of $\nu'(T)$ found in the framework  of
 the one-fluid model of eddy-wave crossover at low temperatures and the experiments. Importantly, in the modeling we have used only
 one phenomenological parameter to fit the zero-$T$ limit of $\nu'\sb{exp}(T)$, while the temperature dependence of the latter
 follows from the model without any additional fitting.

\subsubsection{\label{sss:T-to-0} High temperature range $T  \gtrsim  T_\lambda /2 $}

At $T\gtrsim 1\,$K,   Kelvin waves are already fully damped; see \Fig{fig1}a. This means that for these temperatures we can use
the coarse-grained HVBK \Eqs{NSE}. Using the shell model approximation we find
 the temperature dependence of $\nu\sb s(T)$  in the temperature range $T>0.8\,$K as shown in \Fig{fig5} by the solid red
 line. In the intermediate temperature range $0.9\,$K$<T< 1.1 \,$K this line overlaps with the blue solid line, showing the one-fluid results. The reason for this overlap is very simple: for  $0.9\,$K$<T< 1.1 \,$K, the Kelvin waves are already damped (see
 \Fig{fig1}a), and the normal-fluid eddies at scales $k\ell \sim 1$ are still damped. Therefore, in this range both the  one-fluid
 model  and the coarse-grained model describe the physics equally well. Moreover, the effective viscosity $\nu'$ in the one-fluid approximation, suggested in Ref.\cite{VinenNiemela}
and shown as the black dot-dashed line in \Fig{fig5},  gives the same result as the two our  approaches. At
$T<0.9\,$K the coarse-grained (blue and black) results deviate below the one-fluid prediction which also accounts  for
the energy transfer to the Kelvin waves. This results  in a slower decrease of $\nu'(T)$ with temperature, and finally, in the zero temperature limit (i.e. below $0.5\,$K) this predicts the plateau which is fully determined by the bottleneck energy accumulation at the crossover
between the hydrodynamic and the Kelvin wave regimes of the superfluid motions.

For temperatures above $1.1\,$K,  the normal fluid contribution to the two-fluid dynamics becomes important, and the two-fluid
results in \Fig{fig5}  (red solid line)  deviate above the value of $\nu'\sb s(T)$ which is determined by the superfluid
component alone, cf. Ref.\cite{VinenNiemela}.
This is because in this temperature interval the normal-fluid viscosity $\nu\sb n(T)$ (dashed green line) is much higher  than its
superfluid counterpart $\nu'\sb s$. Therefore there exists an energy flux which is induced by the mutual friction from the less damped superfluid component to the normal fluid component. This is seen in \Fig{fig4}, were $R\sb s>1$ and $R\sb n<1$. Thus  {
$\< |\omega\sb s|^2\>$  is suppressed} and correspondingly $\nu'(T)$  deviates above $\nu'\sb s$ (the black dot-dashed line)
that does not account for the energy exchange. Clearly, at  $T\simeq 1.8\,$K  , when there is no energy exchange, $\nu'(T)$ should be
equal to $\nu'\sb s$. Also it is clear that for $T \gtrsim 1.8\,$K, when the energy flows in the opposite direction (from the  normal- to the super-fluid  component), one expects that $\nu'(T)$ should be  smaller   than  $\nu'\sb s$.    All these expectations are confirmed by the results shown in \Fig{fig5}.

In \Fig{fig5} we also show the  results for $\nu'(T)$ of the  Prague counterflow~\cite{Ladik} and co-flow decay~\cite{Ladik1} experiments and those of the Oregon towered grid\cite{stalp,nimiela2005} experiments. These high-$T$ experimental data have a significant scatter for reasons discussed in \Refs{Ladik,Ladik1,Ladik2}.

 Taking into account the scatter of the experimental data, our computed dependence of $\nu'(T)$  agrees reasonably well with
 experiments in various flows
in the entire range of temperatures from $T\to 0$ up to $T\to T_\lambda$.

\section{\label{s:EnSp}  Coarse-grained, two-fluid dynamics of superfluid turbulence}
This Section concentrates on the high-temperature regime, say $T\gtrsim T_\lambda /2$, when the  small-scale motions (Kelvin waves)
are effectively damped and we can restrict ourselves to a coarse-grained description of the superfluid dynamics in the
continuous-media approximation. In most of this temperature range both the normal and the superfluid components play important role and the two-fluid description is required.

\subsection{\label{ss:2fm} Coarse-grained, two-fluid, gradually-damped HVBK equations}

In this subsection we discuss in more details the gradually-damped HVBK model, presented by \Eqs{NSE}.

\subsubsection{\label{sss:friction} Simple closure for the large-scale energy dissipation due to mutual friction}

Originally in HVBK equations the mutual friction force has the form:
\begin{equation}\label{Fns}
\B F\sb {ns}=  \alpha~\hat{\B \omega}\sb s \times[\B \omega\sb s
\times( \B u \sb n -\B u \sb s) ]   + \alpha'   \B \omega\sb s  \times({\B u\sb s} - \B u \sb n)\ .
\end{equation}
Here $\B \omega\sb s \=\nabla\times \B u\sb s $ is the  superfluid
vorticity and the unit vector $\hat{\B \omega}\sb s \=\B \omega\sb s /\omega\sb
s $ is pointing in the direction of the vorticity. The dimensionless phenomenological parameters
$\alpha $ and $\alpha'$    describe  the dissipative and the reactive mutual friction forces  acting on a vortex line as it moves with respect to the normal component.

In our \Eqs{NSE} we used the simplified form\,\eqref{NSEf-a} of the mutual friction
which  accounts for the fact that the
vorticity in developed  turbulence  is  usually dominated by the
smallest eddies in the system, with  the   Kolmogorov viscous  scale  $\eta$  and with the largest
characteristic wave-vector $k_\eta\sim 1/\eta$. These eddies have
the smallest turnover time $\tau_{\eta}$ which is of the order of
their decorrelation time. On the contrary, the main contribution
to the velocity in the equation for the dissipation of the
$k$-eddies with intermediate wave-vectors $k$,  $k \ll k_{\eta}$,
 is dominated by
the $k'$-eddies with $k'\sim k$. Because the turnover time of
these eddies $\tau_{k'}\gg \tau_{\eta}$, we can justify the approximation~\eq{NSEf-a}
by averaging the vorticiy in \Eqs{Fns} during the time intervals of interest
($\tau_{\eta} \ll \tau\ll \tau_{k'}$). Thus the vorticity $\B \omega \sb s$ may be
considered as uncorrelated with the velocities $\B u \sb s, \B u \sb n$, which are the dynamical variables. More detailed analysis \cite{Lvov} shows that the
approximation~\eq{NSEf-a} directly follows from the Kraichnan's Direct Interaction Approximation in the Belinicher-Lvov sweeping-free
representation for the velocity triple-correlations.

\subsubsection{\label{ss:diss}Vinen-Niemela model for the superfluid energy dissipation}

The energy dissipation term involving $\nu'\sb s$ in Eq. \eqref{NSEa}
for the superfluid velocity attempts to take into account
the existence of quantized vortex lines in an essentially classical regime of motion.  An early approach
to this dissipation term\,\cite{nueff} was  based on a picture of a random vortex tangle moving in a quiescent
normal component. With the definition\,\eqref{oT} this picture leads to the simple equations for the effective viscosity\,\cite{Ladik2}:
\begin{equation}\label{nurnd}
\nu'\sb{rnd}= \alpha \rho \sb s \big / \rho\,,
\end{equation}
shown by a black thin dashed line in \Fig{fig5}. One sees that this result is much lower than the experimental data. Moreover,
 the picture of a random vortex tangle moving in a quiescent normal fluids predicts a higher dissipation compared
 to the realistic situation in which the normal and superfluid velocities  are almost locked together as discussed in \Sec{ss:intro-D}.  The missing physics that needs to be considered to resolve this contradiction is that of vortex reconnections.

During reconnections,  sharp angles necessarily appear on the vortex lines, leading to their fast motion. This motion is uncorrelated with the motion of
other vortex lines (except the ones involved in the reconnection) as well as with the (relatively slow) motion of the normal-fluid
component. This leads to  large local energy dissipation events due to the mutual friction, which  smoothes  the vortex lines and
removes  the regions of high curvature appearing after the reconnection events. A detailed analysis of this and related
effects led Vinen and Niemela\,\cite{VinenNiemela} to suggest the effective superfluid (kinematic) viscosity $\nu^\prime\sb s$ which in
our notations reads:
     \begin{equation}\label{nus}
 \nu'\sb{s}= \beta(T)\, \alpha\kappa  \,, \qquad \beta(T) \= s\, \Big (\frac{c_2 \Lambda}{4\pi}\Big)^2\ .
 \end{equation}
 Here, the parameter  $c_2$ relates the vortex line density and the mean-square  curvature\cite{26,KLPP-2014}, and the parameter $s<1$
 accounts for the suppression of effective line density due to  partial polarization of vortex lines  and was roughly estimated  in
 Ref.\cite{VinenNiemela} as 0.6. The temperature dependence of $\nu'\sb s(T)$ estimated with the help of  Tab.\,III of \Ref{VinenNiemela}
 is presented in Tab.\,\ref{t:1}. The resulting plot of $\nu'\sb s(T)$ is shown in \Fig{fig5} by a thick black dot-dashed line.
 Besides the clear and definitely relevant physics underlying \Eq{nus}, it agrees well with the experiments. That is why in our
 analysis we will include effective damping\,\eqref{nus} in our gradually damped HVBK model \Eq{NSEa}.

\subsubsection{\label{ss:truncated-vs-gd} Truncated vs. gradually-damped HVBK models}

A previous model which is referred to as the  ``truncated HVBK model" of superfluid turbulence was suggested in \Ref{22}. The idea was to account for the
strong suppression of Kelvin waves at high temperatures by simply truncating the  HVBK equation for the superfluid at a cutoff wavenumber $k_\ell  = \tilde \beta /\ell $ (at which the normal fluid is expected to be well damped by  the viscosity), using a  fitting
parameter $\tilde \beta$ of the order of unity.  An obvious limitation of this model is the abruptness of the truncation. An attempt to use this model for the calculation  of the effective viscosity\,\cite{Ladik2} shows that the resulting $\nu'$ [denoted in \,Ref.{\cite{Ladik2} as $\nu\sb{eff}$] may vary  by a factor of about five, when $\~\beta$ changes by the same factor (see Fig.~4, Top,  in \,Ref.\cite{Ladik2}).
Moreover, due to the strong temperature dependence of the normal-fluid viscosity $\nu\sb n$, the truncation scale depends strongly on temperature, as illustrated in our \Fig{fig1}b. This means that  the  fitting parameter $\tilde \beta$ should be temperature
dependent. If so, the truncated HVBK model looses its predictive power. We remark however that the experimental values of the effective viscosity presented in \,Ref.\cite{Ladik2}   unaffected by issues with the truncated HVBK equations discussed here.

 Unfortunately, this is not the only problem in the analysis of $\nu\sb{eff}$ in \Ref{Ladik2}. For interpreting the results of the
numerical simulations of the truncated HVBK model the authors of \Ref{Ladik2} use \Eq{nurnd} for $\nu'\sb{rnd}$ which presumes the random vortex tangle. The values of $\nu'\sb{rnd}$ are shown in \Fig{fig5} by a black thin dashed line. As we already noticed in \Sec{ss:diss}
these  values  are smaller by one order of magnitude  as  compared to the experimental points.
The physical reason for this discrepancy is very simple. The truncated HVBK model ignores the energy dissipation in the reconnection
 events, that, according to our simulations of gradually-damped HVBK model  illustrated in \Fig{fig4}, constitute  more than $90\%$
 of the total energy dissipation in the system.

\subsection{\label{ss:2fluid} Two-fluid Sabra shell-model of turbulent $^4$He}
\subsubsection{\label{sss:Sabra} Sabra-shell model equations}
Following\,\cite{s5} we can present gradually damped HVBK \Eqs{NSE} for isotropic space-homogeneous turbulence as the  system of two shell model equations for the normal $v\sp n\sb m$ and superfluid $v\sp s\sb m$ shell velocities coupled by the friction force term.  In the dimensionless form it may be written as follows:
 \begin{subequations}\label{SM}
 \begin{eqnarray}\label{SMn}
   \Big[\frac{d}{d\tau}+\~\nu \sb n \,  k_m^2  \Big]v\sp {n} _m& =&\mbox{NL}_m\{v\sp {n} _{m'}\}+   \frac{\rho\sb s}{\rho_n} F_m  +f_m\sp
   {n} \, ,\\ \label{SMs}
 \Big[\frac{d}{d\tau}+\~\nu \sb s' \, k_m^2 \Big]v\sp {s} _m  & =&\mbox{NL}_m\{v\sp {s} _{m'}\}-F_{ m}  +f_m\sp {s}\, ,\\\label{t-nu}
 \~\nu\sb n=  \frac{\nu \sb n}{\kappa\,\mbox{Re}_\kappa}\,, &&  \~\nu\sb s' =  \frac{\nu \sb s'}{\kappa\,\mbox{Re}_\kappa}\,, \\ \nn
\mbox{Re}_\kappa=\frac{L\, U\Sb T}{\kappa}\,, &&  U\Sb T^2= 2 \frac{\big [\rho \sb s K\sb s + \rho \sb n K \sb n \big ] }\rho\,,
   \\  \label{omega}
  K\sb s  =  \frac 12 \sum_m   |v_m \sp s|^2\,,&&  \quad K\sb n  =  \frac 12  \sum_m   |v_m\sp n |^2\,,\label{cons}\\ \label{omega1}
F_m=  \alpha \,\omega\sb s(v\sp {s}_m-v\sp {n}_m)\,, \hskip -1 cm &&  \qquad \omega\sb {s,n}^2\equiv   \sum_m k_m^2 |v_m\sp {s,n}|^2
.
\end{eqnarray}
 \end{subequations}
 Here  NL$_m\{v _{m'}\}$ is the Sabra nonlinear term, given by \Eq{Sabra}.
 The dimensionless shell wave numbers $k_m$ are chosen as a geometric
progression $k_m = k_0 2^m$, where~$m=1,2,\dots\,M $ are the shell indices,
 and the dimensionless reference shell wave number is normalized by the inverse outer
 scale of turbulence $1/L$ (to be specific, in our simulations we have chosen $k_0=1/16$).

 The   NL$_m\{v _{m'}\}$   term\,\eqref{Sabra}   conserves
  the kinetic energy (per unite mass) $K\sb s$ and  $K\sb n$ \eqref{cons} provided  $a+b+c=0$ (which is our choice: $b=c=-a/2, ~a=1$).
\
The shell energies $|v_m\sp{n,s}|^2$ correspond to the normal- and the super-fluid energy spectra $\C
E\sb{n,s}(k)$  as follows:
\begin{equation}\label{rel2}
|v_m\sp{n,s}|^2 = k_m \C E\sb{n,s}(k_m)\,, \quad k_m =k_0 \lambda^m\ .
\end{equation}
Here, the factor $k_m$  originates from the Jacobian of transformation from $dk$ to $d\ln k= (1/k) dk $ in the integrals for  the total
energy.   In particular, the  KO-41  spectra with $\C E\sb{n,s}(k)\propto k^{-5/3}$ correspond  to  $|v_m\sp{n,s}|^2\propto k^{-2/3}$
spectra  for the shell  energies.

\subsubsection{Choice of parameters and numerical procedure}
A random $\delta$-correlated in time ``energy-only''
forcing\,~\cite{Sabra},  $f_m\sp s$ and $f_m\sp n$,    was added to the first two shells in   equations for  both the  normal- and the super-fluid components\Eqs{SMn} and \eq{SMs} respectively. Its amplitude was
chosen such that the  total dimensionless energy $K\sb s + K\sb n \simeq 1.$  The standard relation between physical velocity fields
$\B u\sb n(\B r)$,   $\B u\sb s(\B r)$ and shell velocities $v_m\sp n$, $v_m\sp s$ is  a bit involved and will not be displayed here.
Note only that the dimensionless time $\tau$ in \Eqs{SM} is normalized by the turnover time of the energy-containing eddies. What is
more important here is the normalization of the viscosities by $\kappa$ and Re$_\kappa$; see  \Eqs{SM}. The Reynolds number
Re$_\kappa$ is a free parameter of the simulations that determines the width  of the inertial interval:
$\lambda^M\propto\,$Re$_\kappa^{3/4}$.

 The mean effective viscosity in the superfluid subsystem, $\< \nu'\>$, is  calculated from the mean superfluid energy flux,
 $\<\varepsilon\sb s\>$, and the mean enstrophy, $\< \omega\sb s^2 \>$. According to definition\,\eq{oT} and \Eqs{SM}, we have:
\begin{subequations}\label{res-nu}
\begin{equation}\label{res-nu-s}
  \frac{ \< \nu'  \> } {\kappa} =\mbox{Re}_\kappa\frac{\<\varepsilon\sb s\>}{\< \omega\sb s ^2\> }\ .
\end{equation}
Here,    $\omega \sb{s}$ is  given by \Eqs{omega1}, 
and the energy fluxes
 through shell $m$, $\varepsilon\sb{s,n}(k_m)$, are as follows,
\begin{eqnarray}\label{S-flux}
\varepsilon\sb{s,n}(k_m)&=& k_m\, \mbox{Im}[  a\lambda S\sp{s,n}_3(m+1)-c\, S\sp{s,n}_3(m)]\\
 S\sp{s,n}_3(m)&=& v_{m-1}\sp{s,n}v_{m}\sp{s,n} {v_{m+1}\sp{s,n}}^* .
\end{eqnarray}

\end{subequations}
 \Eqs{SM} were solved using the 4th order Runge-Kutta method with an exponential time differentiation  \cite{CoxMatthews-2002}.

 The shell velocities $v_m\sp{s,n}$ were initiated to have the amplitudes proportional to $k_m\sp{1/3}$ and random phases.
The simulations were carried out for temperatures from $T=0.8 \, $K to $T=2.15\,$K using  Re$_\kappa=10^{8}$ and $N=32$ shells.
All other parameters are given in Table \ref{t:1}.
All observables were obtained by  averaging over  about 500 large eddy turnover times. The mean energy fluxes
$\varepsilon\sb{s,n}(k_m)$ were calculated by additional averaging over shells $5\div 13$.

The results of these simulations are shown in \Figs{fig1}b, \ref{fig2}b, \ref{fig3}, \ref{fig4} and \ref{fig5} and were discussed in  \Sec{sss:LT-1fluid}.

\section{\label{s:1-fluid}Low-temperature, one-fluid statistics of superfluid turbulence}
In \Sec{sss:LT-1fluid} we presented an overview of one-fluid description of superfluid turbulence, based on the differential approximation for the energy flux in terms of the energy spectrum itself.
In this section we discuss this differential closure procedure in much more details and derive second-order ordinary differential equation for the superfluid energy spectra $\C E\sb s(T)$   in the entire range of scales, but for $T \lesssim T\sb c/2$.
 \subsection{\label{ss:E-HD} Differential approximation for the energy fluxes of hydrodynamic and Kelvin wave motions}

In this subsection we discuss the analytic form of  the energy flux
 $\varepsilon(k)$ from small $k$ motions toward  the largest  possible  $k$ motions  presenting  an overview of
 results for $\varepsilon(k)$ obtained in  Refs.\,\cite{LNR-2,LN-09,KW-2,KW-T} and  required for further  developments. In
 \Sec{sss:KWS} we begin with the analysis of the expression for $\varepsilon(k)$ for the Kelvin wave region,  $k\ell\ll 1$, in terms of
 its energy spectra $\C E\Sb{KW}(k)$ and then, in \Sec{sss:EHD}, we discuss the expression for $\varepsilon(k)$ for large scale
 hydrodynamic motions in terms of  the hydrodynamic energy spectra.
\subsubsection{\label{sss:KWS} Small scale   motions of Kelvin waves}
It is now recognized that the typical turbulent state of a
superfluid consists of a complex tangle of quantized vortex
lines\,\cite{26} swept by  the velocity field produced by the entire tangle according to the Biot-Savart equation\,\cite{1}. Motions
of the superfluid component with  characteristic scales $R\ll \ell$  may be considered as motions of individual vortex lines, i.e.
Kelvin waves. An important step in studying Kelvin-wave turbulence
was done by Sonin\,\cite{29} and later by Svistunov\,\cite{30}, who found
a Hamiltonian form of the Biot-Savart Equation  for a slightly perturbed straight
vortex line. The final form of this Hamiltonian,  found in Ref.\,\cite{10LLNR} served as a basis for consistent statistical description of
Kevin wave turbulence by Lvov and Nazarenko\,\cite{LN-09} in the framework of  standard kinetic equations for weak wave
turbulence\,\cite{92ZLF,11Nazar}.  This approach~\cite{LN-09,KW-2,KW-T} led to  the spectrum of Kelvin waves
$E_{_{\rm KW}}(k)$.  Below we present a brief overview of these and other pertinent results.

 \paragraph{\label{p:T=0}Zero temperature limit.}  The total ``line-energy density" of Kelvin waves  $E_{_{\rm KW}}  $ (per unit length of the vortex line and normalized by   the superfluid density)   is given by the $k$-integral of  the   energy spectrum  $E_{_{\rm KW}}(k
 )$:
\begin{subequations}\label{defs}
\begin{eqnarray} \label{defsA}
E_{_{\rm KW}} &=&\int   E_{_{\rm KW}}(k ) \, d  k \,, \ E_{_{\rm KW}}(k ) =2 \, \omega(k)\, n(k )\ . ~~~
 \end{eqnarray}
Here $n(k )$ is the wave action, in the classical limit related  to the occupation numbers $N_k$ as follows: $n(k )/\hbar\to N(k )$;
$\omega(k)$ is the frequency of Kelvin waves. For our purposes it is sufficient to use Local-Induction-Approximation (LIA)\cite{26} for
$\omega(k)$:
\begin{eqnarray} \label{defsB}
  \omega(k)= \frac{ \Lambda \kappa }{4\pi}\, k^2\,, \
 \Lambda \equiv  \ln \Big(\frac \ell a_0 \Big)\,,
\end{eqnarray}
\end{subequations}

A previous model of   gradual eddy-wave crossover\cite{LNR-1,LNR-2} was based on  the Kozik-Svistunov (KS) spectrum of Kelvin
  wave turbulence~\cite{KS04}
\begin{equation} \label{KS-spectrum}
E_{_{\rm KW}}\Sp{KS}(k)= C_{_{\rm KS}}\frac{\Lambda\, \kappa^{7/5}\, \epsilon^{1/5}}{ k^{7/5}}\,, \quad \mbox{KS-spectrum .}
\end{equation}
Here $C_{_{\rm KS}}\sim 1$ is a dimensionless constant and $\epsilon$ is the   flux of $E_{_{\rm KW}}(k)$ in the one-dimensional
$k$-space. The KS spectrum~\eqref{KS-spectrum} was obtained in the framework of the kinetic equation \cite{92ZLF,11Nazar}   for weakly
interacting   Kelvin waves under the  crucial assumptions that the  energy transfer in  Kelvin wave turbulence is a
step-by-step cascade, in which only Kelvin waves of similar wave numbers effectively  interact with each other. However in
Ref.~\cite{10LLNR} it was shown that this  locality assumption is not satisfied.  This means that KS-spectrum  is NOT a solution of
the kinetic equations and thus physically irrelevant.

The Kelvin wave turbulence theory was corrected  and a
 new \em local \em  Kelvin wave spectrum was derived by  L'vov and Nazarenko  (LN) in Ref.~\onlinecite{LN-09}:
\be\label{LNspecA}
E\Sb{KW} (k)  = \frac{ C\Sb {LN}}{\Psi^{2/3}}  \,   \frac{\L \, \k \, \epsilon^{1/3}}{   k^{5/3}} \,, \quad \mbox{LN-spectrum}.
\end{equation}
Here $C\Sb{LN}\approx 0.304$\,\cite{KW-2} and the dimensionless
constant  $\Psi$ is given by \Eq{int-psi}.

This KS vs LN controversy triggered an intensive debate (see e.g.
Refs~\cite{LN-debate2,LN-debate3,KS-debate,Sonin,LN-Sonin}),
which is outside the scope of this article. The two predicted exponents, $\frac 75=1.4$ and $\frac53\approx 1.67$ are very close to
each other;
indeed vortex-filament  simulations~\cite{Carlo4} could not distinguish them
(probably because in this numerical experiment   the regime of weak turbulence
on which  the theory is based and which requires a small ratio of the amplitude of the waves compared to the wavelength,
was not the sufficiently realized).
Nevertheless,  more recent   simulations by Krstulovic~\cite{Krs}, based on the
long time integration of   the Gross-Pitaevskii equations and averaged over an ensemble of
initial conditions (slightly deviating from a straight line),  support the LN spectrum. The
most recent vortex-filament  simulations by  Baggaley and Laurie~\cite{BL} observe a remarkable agreement with the LN spectrum with
$C\sp{num}\Sb{LN}\approx 0.308$ close to  $C\sp{anal}\Sb{LN}\approx 0.304$ while $C\sp{num}\Sb{KS} \approx 0.009$  differs from the
KS-estimate $C\Sb{KS}\sim 1$.   Based on these results we will   use LN-spectrum~\eqref{LNspec} in further discussions of the
bottleneck effect.

\paragraph{\label{p:supr} Differential approximation for the Kelvin-wave energy flux.}
 In \Ref{KW-T}, the LN spectrum of Kelvin waves~\eqref{LNspec} allowed to formulate a differential approximation for the energy
 flux,
\begin{subequations}\label{KW-T}
\begin{equation}\label{eps-KW}
    \epsilon\Sb{KW}(k)=-\frac{\Psi^2k^6}{5 (C\Sb {LN}\Lambda \kappa)^3}\, \frac {\partial E\Sb{KW}^3(k)}{\partial k}\,,
    \end{equation}
    which is an important ingredient of the low-temperature, one-fluid differential model.
It was constructed by analogy with \Eq{Leith-n} such as to reproduce the LN spectrum~\eqref{LNspec} together with  the
thermodynamical equilibrium solution $E\Sb{KW}(k)=$const.  The  approximation~\eqref{eps-KW}  plays an important role in the
discussion of the temperature dependence of the effective superfluid viscosity~$\nu'(T)$.

  To analyze the temperature suppression of the Kelvin waves energy spectrum, consider the energy balance equation
\begin{equation}\label{KWbal}
\frac{d \epsilon\Sb{KW}(k)}{d k}=- \frac{\alpha  \Lambda}{4\, \pi}\, \kappa k^2 E\Sb{KW}(k)\,,
\end{equation}
\end{subequations}
whose right-hand-side accounts for the dissipation of the Kelvin waves in the simplest form suggested by Vinen in Ref.\,\cite{KWdiss}.  The
approximate solution of \Eqs{KW-T},  found in Ref.~\cite{KW-T},  is as follows:
\begin{subequations}\label{sol-KW}
\begin{eqnarray}\label{KW-Ea}
 E\Sb{KW}(k,T)&\approx& \frac{ C\Sb {LN}}{\Psi^{2/3}}  \,   \frac{\L \, \k \, \epsilon_0^{1/3}}{   k^{5/3}} \Big [ 1 - \Big ( \frac k
 {k\sb{max} }\Big )^{4/3} \Big ]\,, \\ \nn &&  1/\ell \leqslant k \leqslant k\sb{max}\ .
 \end{eqnarray}
Here $\epsilon_0\equiv \epsilon\Sb{KW}(1/\ell)$ is the energy influx into system of KWs at $k\sim 1/\ell$.  Notice that  all the
temperature dependence of $E\Sb{KW}(k,T)$ is absorbed in
 $k\sb{max}(T)$ given by:
\begin{equation}\label{KW-Eb}
 k\sb{max}=k\sb{max}(T)\approx \frac {14 \sqrt {\Psi \epsilon _0}}{\Big [ \sqrt {\alpha (T)C\Sb {LN}}\Lambda \kappa\Big ]^{3/2}} \ .
 \end{equation}
\end{subequations}
The analytical solution~\eqref{sol-KW} is in qualitative agreement with the numerical results shown in \Fig{fig1}a.

\subsubsection{\label{sss:EHD} Large-scale hydrodynamic region}
In the hydrodynamic range of scales, the Biot-Savart description of superfluid turbulence is too detailed for our purposes and we
can return to the continuous  medium approximation, \Eqs{NSE}, used in the high temperature regime. The difference with \Sec{s:EnSp}
is that now we will first perform the statistical averaging of the velocity field, and only then analyze the resulting equations for
the energy spectra in the one-fluid approximation. This different strategy is dictated by a natural requirement that Kelvin waves and
hydrodynamic eddies have to be treated in a similar formal scheme in order to describe the intermediate region of scales where one
type of motion continuously turns  into the other. As a candidate for this scheme we choose a differential closure that allows us to
express the energy flux as a differential form of the energy spectra. For the Kelvin waves this approximation was given by
\Eq{eps-KW} and for hydrodynamic eddies it is discussed below.

  The simplest approximation for the hydrodynamic energy flux $\varepsilon\Sb{HD}(k)$, based on the Kolmogorov  idea of the locality
  of the energy transfer and dimensional reasoning goes back to Kovasznay 1947 paper~\cite{Kov47}:
 \label{flux}
\begin{equation}\label{fluxA} \varepsilon\Sb {HD} (k) \simeq C  \C [\C E\Sb {HD}(k) ]^{3/2} k^{5/2}\ .
\end{equation}
Here  $C\sim 1$ is a  dimensionless constant.
 The basic idea of such models is that the nonlinear terms, being of the
simplest possible form, should preserve the original turbulence
scalings and, in particular, predict correctly the Kolmogorov
cascade. Indeed, in the stationary case and in the
absence of dissipation the energy flux becomes  $k$-independent,   $\varepsilon\Sb {HD} (k)\Rightarrow \varepsilon\Sb {HD} $.  Then
\Eq{fluxA}  turns into  the
Kolmogorov-Obukhov $5/3$--law for $\C E\Sb{HD}(k)$, given by \Eq{V6-1}.

Unfortunately, the simple relation~\eq{fluxA} does not describe the thermodynamic equilibrium (with equipartition of energy between
states), when the energy flux vanishes for  $\C E\Sb{HD} \propto k^2$. This disadvantage is corrected in the Leith-1967   differential
model~\cite{Leith67}, given by \Eq{Leith-n}.
 This approximation coincides  dimensionally  with the   Kovasznay model~\eq{fluxA}, but has a derivative $d [\C
E\Sb{HD}(k)/k^2]$ that guarantees that $\varepsilon\Sb{HD} (k)=0$, if $\C E\Sb{HD}(k) \propto k^2$.
The numerical factor  $\frac18$, suggested in \cite{Nazar-Leith},   gives the value of the Kolmogorov constant $C\Sb {K41}=(24/11)^{2/3}\approx 1.7 $ in \Eq{V6-1} that is reasonably close to the experimentally observed value.

A generic hydrodynamic spectrum with a constant energy flux was found in \cite{Nazar-Leith} as a solution to the equation ~$\varepsilon\Sb{HD}
(k) = \varepsilon\Sb{HD} = \mathrm{const}$: 
\begin{eqnarray}
  \label{K41}
  \C E\Sb{HD}(k) = C\Sb{K41} \frac{\varepsilon\Sb{HD}^{2/3}} {k^{5/3}} \left[ 1+\Big ( \frac{k}{k\sb{eq}}\Big)^{11/2}
  \right]^{2/3}\,,
\end{eqnarray} 
 in which $k\sb{eq}$ is as yet a free parameter describing the crossover  between  the low-$k$ KO41 spectrum~\eq{V6-1} and  the thermalized  part of the spectrum, $\C E\Sb{HD}(k)\propto k^2$ with equipartition of energy at large $k$.

Notice, that Eq.\,\eqref{K41} does not account for the energy dissipation due to the mutual friction and the viscosity. We will do this
later, introducing dissipation terms in the energy balance equation and numerically solving them.


\subsection{\label{s:BB} One-fluid differential model of gradual   eddy-wave crossover}

 \subsubsection{``Line-" and ``volume-" energy densities, spectra and fluxes }
  Our goal here is to formulate a model which will allow to  describe in a unified form the hydrodynamic energy flux $\varepsilon
  \Sb{HD}(k)$ at small $k$ and the corresponding objects for the Kelvin waves. However we cannot do it straightforwardly using
  the equations for  $\epsilon \Sb{KW}(k)$ and the spectrum $\C E \Sb{KW}(k)$. The reason is simple: the hydrodynamic and Kevin wave objects have different physical meaning and different dimensions. Indeed, the hydrodynamic motions fill the three-dimensional space
  (volume), their  energy density $\C E\Sb{HD}$ per unit mass has a dimension $[\C E\Sb{HD}]$=cm$^2$/sec$^2$. Accordingly, the
  dimensions of energy spectrum $\C E \Sb{HD}(k)$ and the energy flux $\varepsilon \Sb{HD}(k)$   are as follows:
\begin{subequations}\label{dim}
\begin{equation}\label{HD-dimA}
[\C E \Sb{HD}(k)]=\mbox{cm}^3/\mbox{sec}^2\,, \quad [\varepsilon \Sb{HD}(k)]=\mbox{cm}^2/\mbox{sec}^3\ .
\end{equation}
 On the other hand, Kelvin waves propagate along one-dimensional lines -- vortex filaments. Therefore the  energy density of Kelvin
 waves on individual vortex filament $E\Sb{KW}$ is normalized by unit vortex length and (for the sake of convenience) by superfluid
 density. Therefore its dimension  is $[E\Sb{KW}]$=cm$^4$/sec$^2$.  Then the dimensions of the corresponding energy spectrum and the energy flux are:
  \begin{equation}\label{KW-dimA}
[  E \Sb{KW}(k)]=\mbox{cm}^5/\mbox{sec}^2\,, \quad [\epsilon \Sb{KW}(k)]=\mbox{cm}^4/\mbox{sec}^3\ .
\end{equation}
\end{subequations}

Different normalization of the same objects dictates the relation between them
    in a statistically homogeneous and isotropic vortex tangle with the line density $L=1/\ell^2$
    \begin{equation}\label{rel1}
 {\cal E}_{_{\rm KW}}(k)  = \frac{E_{_{\rm KW}}(k)}{\ell^2}\, , \   \varepsilon_{_{\rm KW}} = \frac{\epsilon_{_{\rm KW}}
 }{\ell^2} \  .
\end{equation}

\subsubsection{\label{sss:blend}Energy balance equation}
Consider a general form \eq{sBal} of the continuity equation for the energy density $\C E\sb s(k,t)$ of the  isotropic space-homogeneous
turbulence of the superfluid component
which accounts for the energy dissipation with the help of the Vinen-Niemella viscosity.
 The remaining  physical problem here  is how to describe the energy density  $\C E\sb s(k,t)$,  the energy flux over scales
 $\varepsilon\sb s (k,t)$, and the damping term $\C D(k,t)$ in the entire range of wave vectors $k$, including the  intervortex
 scales  $k\sim 1/\ell$.  A  step toward this direction was suggested in Ref.~\onlinecite{LNR-1} in the form of the ``Eddy-wave model" in which the superfluid motions with scales $R \sim \ell$ are considered as a superposition of two coexisting and  interacting   types of motion: random eddies and Kelvin waves.  In some sense the problem here is similar to the description of the mechanics of the matter
 at  intermediate range of scales, where it behaves like   particles and waves simultaneously.  In quantum mechanics it was suggested
 to formulate explicitly  the basic equation of motion (the Schr\"{o}dinger equation) and to compare its prediction with
 observations. We are not so ambitious, our goal is to discuss below  a set of uncontrolled approximations, based currently only on our
 physical intuition, which  leads to an explicit set of model equations  having a predictive power. As we will see below,  our model
 predicts that in the range $R\sim \ell$, due to the bottleneck energy accumulation,  the energy distribution between scales is
 close to the energy equipartition, like  in the thermodynamic equilibrium. It is well known that in  thermodynamic equilibrium
 the statistics is universal and independent of the details of interaction. Therefore we hope that many details of the vortex
 dynamics (including the vortex reconnections), that are ignored in our model,  do not affect the model results: we believe that
 these results are closer to the reality than the model itself.

\subsubsection{Gradual model for the energy spectra of superfluid component}

 The basic physical idea is to approximate the total turbulent superfluid energy density  $\C E\sb s(k,t)$ as a sum of the
 hydrodynamic energy  spectrum  $\C E\Sb{HD}\sp s(k,t)$
 and the energy  spectrum  of the Kelvin waves $\C E\Sb{KW}(k,t)$, with
 the energy distribution between the  components  depending  only on the dimensionless blending function $g(k\ell$) of the dimensionless wave-number $k\ell$:
 \begin{eqnarray} \nn  \C E\sb s (k,t)&=&\C E\Sb{HD}\sp s(k,t)+\C E\Sb{KW}(k,t)\,,  \\
 \C E\Sb{HD}\sp s(k,t) &=& g(k\ell )  \, \C E \sb s(k,t)\,, \label{mod3} \\
  \C E\Sb{KW}(k,t) &=&[1- g(k\ell)]\,   \C E\sb s  (k,t)\ .  \nn
 \end{eqnarray}
In order to find a qualitative form of the blending function $g(k\ell)$ we follow Ref~\cite{LNR-1}.  Consider a system of locally near-parallel vortex lines (in
the vicinity of some point $\B r_0$), separated by the mean distance $\ell$.  Denote the individual vortex lines by an index $j$.
Notice that in principle the same vortex line can go far away and
come close to $\B r_0$ several times. To avoid this problem one should assign the same vortex
line a different index $j$ if it leaves (or enters) the ball of radius $  \ell\sqrt \L$
centered at $\B r_0$. Each vortex line produces a superfluid velocity field $\B u_j\sp s(\B r)$, which can be found by the
Biot-Savart Law.

 The total superfluid kinetic energy  density (per unit mass) $\C E\sb s =\frac12 \sum_{i,j} \<\B u_i\sp s\cdot \B u_j\sp s\> $ may be divided into two parts,
 $E\sb s =\C E_{1\rm  s}+\C E_{2\rm  s}$, where
\begin{eqnarray}\label{eq1}   \C E_{1\rm  s}&\=& \frac 12 \sum_j \< (u_j\sp s)^2\>\,,  \\ \nn
\C E_{2\rm  s}&\=&  \frac 12 \sum_{i\ne j}\<
\B u_i\sp s\cdot \B u_j\sp s\> = \sum_{i< j}\< \B u_i\sp s\cdot \B u_j\sp s\>\ .
\end{eqnarray}
The same subdivision can also be made  for the energy spectrum in the (one-dimensional) $k$-space,  $\C E\sb s(k) = \C E_{1\rm s}(k)
+\C E_{2\rm s}(k)$, with two terms, that may be found via $\B k$-Fourier components of the superfluid velocity fields $\B v_j\sp s(\B
k)$ similar to \Eq{eq1}.
Now the idea is as follows: the energy $\C E_{1\rm s}(k)$ is defined  by the form of the individual vortex lines that is determined
by the Kelvin waves, while the energy  $\C E_{2\rm s}(k)$ depends on correlations in the form of different vortices, that produce
collective, hydrodynamic type of motions.  Therefore $\C E_{1\rm s}(k)$ may be associated with the Kelvin waves energy, $\C E_{1\rm
s}(k)\Rightarrow \C E\Sb{KW}(k)$, while  $\C E_{2\rm s}(k)$  has to be associated with the superfluid
hydrodynamic energy,  $\C E_{2\rm s}(k)\Rightarrow \C E\Sb{HD}\sp s(k)$. This allows us to conclude that
\begin{equation}\label{eq2}
g(k\ell)= \big[1 + \C E_{1\rm s}(k)/\C E_{2\rm s}(k)\big]^{-1}\ .
\end{equation}
The rest are technical details  presented in  Ref.~\onlinecite{LNR-1}, where it was concluded that in practical calculations it is
reasonable to  use an analytical form $g(k\ell)$ of the blending function
\begin{eqnarray}\label{est5}
  g(k\ell)  &=&  g_0\big[c_1\,\ln (\L+7.5)\ k\ell\big]\,,\\ \nn
  g_0(k\ell)& =& \Big[ 1+ \frac{(k\ell)^2\exp(  k\ell )}{4\pi (1+  k\ell)}\,\Big]^{-1}\,,
\end{eqnarray}
where $c_1\approx 0.32$ is the fitting parameter, chosen in~\cite{LNR-1}.

\subsubsection{\label{sss:flux}Gradual model for the superfluids   energy flux $\ve\sb s(k,t)$}
 Modeling the total superfluid energy flux over scales, $\varepsilon\sb s(k,t)$, is less straightforward than the model~\eq{mod3}
 for the energy itself, $\C E\sb s(k,t)$. We first  assume that  $\ve\sb s  (k,t)$ may be presented as the sum of the fluxes over
 hydrodynamic and Kelvin wave components,
 \begin{subequations}\label{mod5}
 \begin{equation}\label{mod5a} \ve \sb s= \~ \ve\sp s\Sb {HD}+ \~\ve \Sb{KW}\,,
  \end{equation}
  however the fluxes $\~ \ve\sp s\Sb {HD}$ and  $\~\ve \Sb{KW}$ are \emph{not equal} to the fluxes $\ve\sp s \Sb {HD}$, \Eq{Leith-n},
  and $\ve \Sb{KW}$ in \emph{ isolated} hydrodynamic and Kelvin wave systems.  Equation  \eqref{eps-KW} for $\epsilon\Sb{KW}(k)$
  in the volume normalization~\eqref{rel1}  {takes the form\,\eqref{vareps-KW}}.

 The fluxes $\~ \ve\Sb {HD}\sp s$ and  $\~\ve \Sb{KW}$    contain  additional cross-contributions   $\ve\Sb{HD } \Sp{KW}(k)$  and
   $\ve\Sb{KW } \Sp{HD}(k)$ that originate from the interaction  of  two types of motion, hydrodynamic and Kelvin waves:
 \begin{equation}\label{mod5b}
    \~ \ve\Sb {HD}\sp s=  \ve\Sb {HD} + \ve\Sb {HD}\Sp{KW}\,, \quad
    \~ \ve\Sb {KW}=   \ve\Sb {KW} + \ve\Sp {HD}\Sb{KW}\ .
     \end{equation}
  We  modeled the cross-terms in the  linear approximation with respect to the energies (i.e. the Hydrodynamic energy affecting
  the Kelvin waves flux and vice versa):
 \begin{eqnarray}\label{mod5c}
 \ve\Sb{HD } \Sp{KW}(k) =\C F\Sb {HD} \{\C E\Sb {HD }\sp s\} \  d \, [\C E\Sb {KW}(k)/ k\sb{en}^2] /d k^2 \,, \\ \nn
 \ve \Sb{KW}\Sp{HD}(k)  =\C F\Sb {KW} \{\C E\Sb {KW}\} \ d \,  [\C E\Sb {HD}\sp s(k)/k^2] /d k^2\,,
 \end{eqnarray}
 where $  k\sb{en}$  is the wave number, at which $g(k\sb {en}\ell)=\frac 12$.
The differential form of these contributions follows from a physical hypothesis that these terms should disappear (or become negligibly
 small) when the influencing subsystem is in thermodynamical equilibrium, i.e. when $\C E\Sb {HD}\sp s\propto k^2$ and  $\C E\Sb
 {KW}\propto k^0=$const. Functionals of the corresponding energies,  $\C F_{\dots}\{\dots\}$ may be modeled  by dimensional
 reasoning, in the same way as Eqs.~\eqref{Leith-n} and  \eqref{eps-KW} were formulated for the fluxes. The resulting equations for
 $\C F\Sb{HD}$ and  $\C F\Sb{KW}$ may be written in the form:
 \begin{eqnarray}\label{mod5d}
 \C F\Sb {HD} \{\C E\Sb {HD }\sp s\} &=& C\Sb{HD}\, \sqrt{k^{11} \C E\Sb{HD}\sp s(k)}\,, \\ \nn  \C F\Sb {KW} \{\C E\Sb {KW}\} &=&
 C\Sb{KW}(k\ell)\, k^{2}\sb {en}\C E\Sb{KW}^4(k)\kappa^{-7}\ .
 \end{eqnarray}
Here
 \begin{equation}\label{mod5e}
C\Sb{HD}=- 1/ 8\,, \quad C\Sb{KW}(k\ell)= - 5 (k\ell)^8/7 \Lambda ^5\ .
\end{equation}
 \end{subequations}
as  explained in Ref.~\cite{LNR-1}.  The resulting model for the total energy flux  $\ve(k)$  follows from \Eqs{mod5}:
 \begin{eqnarray}\nn
  \ve\sb s(k)&=& - \Big\{ \frac 18 \sqrt{k^{11} g(k\ell)\C E \sb s(k)} \\ \label{mod4}  
  &&    + \frac{3}{5}\frac{\Psi^2\,(k\ell)^6k\sb {en}^2  [1-g(k\ell)]^2 \C E\sb s(k)^2}{  (C\Sb{LN}\Lambda\, \kappa  ) ^3 } \,\Big\}
  \\  \nn 
  &&    \times\ \frac{d}{d k}\Big\{ \C E \sb s(k)\Big[ \frac{g (k\ell )}{k^2}+ \frac{1-g (k\ell)}{k\sb{en}^2}\Big] \Big\}\, .
 \end{eqnarray}
Only with the   choice~\eq{mod5e} the resulting \Eq{mod4} for $\ve\sb s(k)$   vanishes in  thermodynamic equilibrium (with $\C
E\sb s(k)\propto k^2$ in the hydrodynamic regime, $k<k\sb{en}$ and with  $\C E \sb s(k)=$const. in the Kelvin waves regime,  $k>k\sb{en}$,)  as required.

\subsubsection{\label{sss:no-dimen} Dimensionless form of the gradual one-fluid model}
The resulting \Eqs{sBal}, \eqref{mod3}, \eqref{est5} and  \eqref{mod4}  represent our  eddy-wave model of superfluid turbulence in the
one-fluid approximation, which neglects motions of the normal-fluid component, assuming $\B u\sb n=0$. Now we introduce dimensionless
variables:
\begin{subequations}\label{dif-m}
\begin{equation}\label{A1}
x=k\ell\,, \quad  e(x)=\frac{\ell\, \C E\sb s (x)}{\kappa^2}\,, \quad  \epsilon(x)=\frac{\ve\ell^4}{\kappa^3}\,
\end{equation}
and express \Eq{mod4} in a dimensionless form which is convenient for numerical analysis:
\begin{eqnarray}\nn
 &&\frac{d}{dx}\Bigg \{ \Big [ \frac18 \sqrt{x^{11}g(x)e(x)}+ \frac35 \frac{\big[\Psi x\sb{en} x^3 (1- g(x)e(x))\big]^2}{(C\Sb
 {LN}\Lambda)^3}\Big ]\\ \label{A2}
 &&\times \frac d{dx}\Big\{ e(x)\Big [ \frac{g(x)}{x^2}+ \frac{1-g(x)}{x\sb{en}^2}\Big ]\Big \}\Bigg\}\\ \nn &=&\alpha\Big \{
 \omega\Sb T g(x)+ x^2\beta(T) \Big \}e(x)\, ,\\
 \label{X2}
  &&x\sb {en}\= k\sb{en}\ell  \simeq    6.64 \big /  \ln (\L+7.5)\ .
\end{eqnarray}
In the dimensionless form there is a constraint on the energy flux:
\begin{equation}\label{A3}
2\int _0^\infty x^2 g(x) e(x) d x=1\ ,
\end{equation}
following from the assumption that the vorticity is dominated by the scales of  order $1/\ell$, given by Eq.\,(31b) in
Ref.\,\cite{LNR-2}.
The function $\Psi$ defined by \Eq{int-psi} in dimensionless variables should be found self-consistently by enforcing the following
 condition,
 \begin{equation}\label{A4}
 \Psi=\frac{8\pi}{\Lambda}\int_{x_{1/2}}^\infty [1-g(x)]e(x) d x\,, \quad g(x_{1/2})=\frac 12\ .
\end{equation}

\end{subequations}

\subsubsection{\label{sss:num} Numerical procedure}
We  solved the  integro-differential \Eq{A2} numerically starting from the large $x$ region.

To formulate two boundary conditions at large $x$, we use an analytical form of the Kelvin waves spectrum. In dimensionless form \Eq{KW-Ea}
reads:
\begin{subequations}\label{A5}\begin{eqnarray}\label{A5a}
e\Sb{KW}(x\to x\sb{max})&=& \frac{C\Sb{LN}\Lambda \epsilon_0^{ 1/3}}{\Psi^{2/3}x^{5/3}}\Big[ 1- \Big ( \frac x {x\sb{max}}\Big)
^{4/3}\Big ] ,~~~~~\\ \label{A5b}
x\sb{max}&\approx& \frac {14 \sqrt {\Psi \epsilon _0}}{(\alpha\, C\Sb {LN}\Lambda^2)^{3/4}}\ .
\end{eqnarray}
\end{subequations}
Now we can take as the boundary conditions the values of $e\Sb{KW}(x)$ at two points, $\epsilon(x\sb{max}- x_1)$ and
$\epsilon(x\sb{max}- x_2)$ with some appropriate values of $x_1$ and $x_2$ (say 5 and 10) for very small $\alpha$. The results of
these simulations are  shown  in \Figs{fig1}a, \ref{fig2}a,  and \ref{fig5} and were discussed in \Sec{s:intro}.


\section{\label{s:Disc} Summary   and   discussion}

In this paper, we  have generalized the zero temperature theory~\cite{LNR-1,LNR-2} of the energy and  the vorticity spectra
in superfluid turbulence  to
non-zero temperatures up to $T\to T_\lambda$, accounting for the effect of the mutual friction and motion of the normal
fluid component. In particular we describe the  influence of the  temperature on the bottleneck energy accumulation near  the inter-vortex scales.\vskip 0.2cm

  {\textbullet~The gradually damped HVBK \Eqs{NSE} include the Vinen-Niemela  superfluid viscosity\,\eqref{nus} with a fitting parameter $s\approx 0.6$ which was chosen in their paper\,\cite{VinenNiemela}. Besides this,  our Sabra-model \Eqs{SM} which is based on \Eqs{NSE} and is used in the $T>T\sb c/2$ range,  has no additional fitting parameters.}

 \textbullet~ The differential one-fluid model of superfluid turbulence  \Eqs{A2}, used in the $T<T\sb c/2$ range, has only one fitting parameter $c_1\approx 0.32$, entering into the blending function~\eqref{est5} and chosen in \Ref{LNR-2}.  Besides this the model has no additional fitting parameters. Thus in the entire approach we used only two fitting parameters which were chosen in previous papers.

 \textbullet~We have shown that for $T \lesssim 0.5\,$K Kelvin waves are excited in the range of scales  from $k\ell \sim 1$ up to some
 temperature dependent cutoff $k\sb{max}$~\eqref{KW-Eb}; see \Fig{fig1}a.   For $k\ell  \gtrsim 20$  Kelvin waves  have the LN-energy
 spectrum~\eqref{LNspec} with a constant energy flux, $\C E\Sb{KW}\sp s (k)\propto k^{-5/3}$, while in the crossover region (about
 one decade around $k\ell\sim 1$) there exists a flux-less spectrum $\C E\Sb{KW}\sp s (k) \approx$const corresponding to
 the thermodynamic equilibrium with  the energy equipartition between Kelvin waves with different $k$. In this temperature range the
 effective superfluid viscosity  may be considered as temperature independent and equal to its zero-temperature limit, $\nu'(T)\approx
 \nu'(0)$. Also, a minor amount of the normal-fluid component may be completely ignored. \vskip 0.2cm

 \textbullet~When $T$ exceeds $\simeq 0.5\,$K,  the constant energy flux range of Kelvin waves disappears and the flux-less range
 (with  $\C E\Sb{KW}\sp s (k) \approx$const) begins to shrink; see \Fig{fig1}a. This leads to the temperature suppression of the
 bottleneck energy accumulation.  As a result,  the superfluid square vorticity, $\< |\omega\sb s|^2\>$, decreases (see \Fig{fig2}a),
 leading  to increase in $\nu'(T)$, in accord with the experimental observations; see \Fig{fig5}. Up to $T\simeq 1.1\,$K, a small
 amount (below 1\%) of the normal-fluid component may be considered as being at rest, at least in the  region $k\ell \sim 1$, which determines the leading contribution to $\< |\omega\sb s|^2\>$.  As a result, both our models (the one-fluid gradual model of the bottleneck
 crossover, which account for the presently negligible energy flux by Kelvin waves, and the gradually damped  HVBK model, that ignores Kelvin
 waves, but accounts for the presently negligible normal-fluid motions) are valid for $0.5\,$K $ \lesssim T \lesssim 1\,$K. The temperature
 dependence of  $\nu'(T)$ predicted by these models (red and blue solid lines in \Fig{fig5}) practically coincide  for $0.5\,$K$
 \lesssim T  \lesssim 1.1\,$K.  \vskip 0.2cm

 \textbullet~In the high temperature regime,   $T \gtrsim T_\lambda /2\simeq 1.1\,$K, the normal fluid component begins to play some
 role in the temperature dependence of $\nu'(T)$.
 In spite of the almost full super- and  normal-fluid velocity locking  (see \Fig{fig3}) there is a significant  energy exchange
 between the components (see \Fig{fig4}), caused by the mutual friction and the velocity decorrelation  near $k\ell\sim 1$. This
 physical effect is important by itself, although it leads only to a small, (but visible)  deviations of the resulting temperature
 dependence of $\nu'(T)$ (solid red line in \Fig{fig5}) from the Vinen-Niemella model\cite{VinenNiemela} of $\nu'(T)=\nu'\sb s(T)$,
 \Eq{nus}, shown by black dot-dashed line.

  \textbullet~ Since there is no detailed information on the superfluid and normal fluid energy spectra, especially for large $k$ and
  low temperatures (see e.g. review~\cite{BLR}),
 we compare our results with the experiments for the temperature dependence of the effective kinematic viscosity
   $\nu'(T)$. The latter  was measured in the temperature range from $0.08\,$K to $2.15\,$K by the Manchester
   spin-down\cite{Manchester-exp} and ion-jet\cite{Golov2} experiments, as well as   the Oregon towed-grid\cite{stalp} and the
   Prague counter-flow\cite{Ladik} experiments, all shown in \Fig{fig5}. Our  computed temperature dependence of the effective
   viscosity $\nu'(T)$  agrees qualitatively with the experimental data in the entire temperature range: from  $T\to 0$ up to  $T\to
   T_\lambda$.  We consider this agreement as a strong evidence that our low-temperature, one-fluid differential model and high
   temperature coarse-grained gradually damped two-fluid  HVBK model capture the relevant basic physics of the turbulent behavior of
   $^4$He.

 \textbullet The models considered in this paper are intended for systems whose anisotropy effects are
  not substantial. Strong external rotation may change the behaviour by
enforcing a strong polarisation of the superfluid vortex bundles leading to the suppression of reconnections. In turn, suppressed reconnections
result in an enhanced bottleneck accumulation of the turbulent spectrum near the crossover scale $\ell$ and, as a result, in a decrease of
the effective viscosity $\nu'$. We leave the study of such an effect of  strong polarisation on the bottleneck phenomenon to future.

  \textbullet~In this  paper, we have ignored the effect of mutual friction
   on the small $k$ region of scales in the low temperature regime, when the normal component is rare and motionless due to a very
   large kinematic viscosity ($\nu\sb n >38\,  \kappa $ for $T\leqslant 1\,$K). This is justified when the range of scales greater
   that $\ell$ is not very wide, as in all existing \He4   experiments. Theoretically, the mutual friction effect grows as $k$ is
   decreased and, if the low-$k$ range is wide, the spectrum would inevitably reach a friction-dominated scaling regime with a
   power-law exponent equal $-3$, see \Ref{LNV}. Such a regime, which is even more natural in $^3$He turbulence, may lead to a vortex
   tangle decay law with $L \sim 1/t$. We leave the study of this dissipative regime for the future.

\section*{Acknowledgements}

We acknowledge the contribution of
Oleksii Rudenko who participated in this project in its preliminary stage.
S.N. gratefully acknowledges support of a grant ``Chaire Senior PALM TurbOndes'' and the hospitality of the SPEC lab, CEA,
Saclay.



\appendix

\section{\label{a:two-f} Simple model of cross-velocity correlations in superfluids}
 Our  goal here  is to suggest a relatively simple and physically transparent model of the  cross-correlation function of the normal
 and superfluid velocities, that leads to \Eq{lim1} in the  simplest case of homogeneous, isotropic turbulence of incompressible
 turbulent motions of $^4$He.

To start, we recall some definitions and relationships, required for our derivation, which are well-known in statistical physics. The
first one is  Fourier transform  in the following normalization:
 \begin{equation}\label{def5}
\B v\sb{n,s}(\B k,t) \=  \int \frac{d \B k }{(2\pi)^3}\, \B
u\sb{n,s}(\B r,t) \exp(-i\B k\* \B r)]\ . \end{equation} 
Next we define simultaneous correlations and cross-correlations in the $\B k$-representation, [proportional to $\d(\B k - \B k')$ in
homogeneous case]:
\begin{subequations}\label{corr}
\begin{eqnarray}\label{corr-nn}\< \B v\sb n(\B k,t)\*\B v^* \sb n(\B k',t)  \>&=& (2\pi)^3 G\sb{nn}(\B k)\, \d(\B k-\B k')\,,~~~ \\
\label{corr-ss}
\< \B v\sb s(\B k,t)\*\B v^* \sb s(\B k',t)  \>&=& (2\pi)^3 G\sb{ss}(\B k)\, \d(\B k-\B k')\,, \\ \label{corr-ns}
\< \B v\sb n(\B k,t)\*\B v^* \sb s(\B k',t)  \>&=& (2\pi)^3 G\sb{ns}(\B k)\, \d(\B k-\B k')\ .
\end{eqnarray}\end{subequations}
It is known that their $\B k$-integration produces  one-point correlations:
\begin{subequations}\label{def6}
\begin{eqnarray}\label{def6a}
 \int \frac{d \B k }{(2\pi)^3}G\sb {nn}(\B k)&=& \< |\B u\sb {n} (\B r, t)|^2 \>\,,\\ \label{def6b}
  \int \frac{d \B k }{(2\pi)^3}G\sb {ss}(\B k)&=& \< |\B u\sb {s} (\B r, t)|^2 \>\,,\\ \label{def6c}
   \int \frac{d \B k }{(2\pi)^3}G\sb {ns}(\B k)&=& \< \B u\sb {n} (\B r, t)\*
   \B u\sb s (\B r, t) \>\ .
\end{eqnarray}\end{subequations}
In isotropic  case, each of three correlations $G_{\dots} (\B k)$ is independent of the direction of $\B k$:  $G_{\dots} (\B
k)=G_{\dots}  (k)$ and $\int \dots d\B k= 4\pi \int \dots k^2 \, dk$. Together with \Eqs{def1c} and \eq{def6} this gives:
 \begin{eqnarray}\nn 
\C E\sb{n}(k)&=& \frac{k^2}{4\pi^2}G\sb{nn}(k)\,, \quad \C E\sb{s}(k)= \frac{k^2}{4\pi^2}G\sb{ss}(k)\,,\\ \label{def1a}
\C E\sb{ns}(k)&\=& \frac{k^2}{4\pi^2}G\sb{ns}(k)\ .
\end{eqnarray}

To begin with the derivation of  \Eq{lim1}, we simplify \Eqs{NSE} for the superfluid and the normal velocities, $\B v\sb s(k,t)$ and $\B v\sb
n (k,t)$, by modeling the nonlinear terms in the spirit of the Langevin  approach, i.e. replacing them  by a sum of respective  damping
terms   $\g\sb s \B v\sb s(k,t)$ or $\g\sb n \B u\sb n(k,t)$  and  random, delta-correlated in time force terms $\B f\sb s (\B k, t)$ or $\B f\sb
n (\B k, t)$ with Gaussian statistics and zero cross-correlations:
 \begin{eqnarray} \nn
\<\B f\sb s (\B k, t)\cdot \B f\sb s^*  (\B k', t') \>&=& (2\pi)^3 \d (\B k -\B k') \delta (t-t')f^2\sb {ss}(\B k) \,, \br
\<\B f\sb n (\B k, t)\cdot \B f\sb n^*  (\B k', t') \>&=& (2\pi)^3 \d (\B k -\B k') \delta (t-t')f^2\sb {nn}(\B k) \,, \\
\label{fdef}
\<\B f\sb s (\B k, t)\cdot \B f\sb n^*  (\B k', t') \> &=& 0\ .
\end{eqnarray}
In the $\B k$-representation, the resulting equations read:
\begin{subequations}\label{NSE2}\begin{eqnarray}\label{NSE2a} 
  \frac{\p  \B v\sb s(\B k, t)}{\p t}+ \G \sb s  \B v\sb s(\B k, t) &\!=\!&\alpha  \bar \o \sb s  \B v\sb n(\B k, t)  + \B f\sb s(\B
  k, t),  ~~~~~~~~~~~
 \\ \label{NSE2b}
 \frac{\p  \B v\sb n(\B k, t)}{\p t}+ \G \sb n  \B v\sb n(\B k, t)  &\!=\!& \frac{\alpha \r\sb s}{\r \sb n} \bar \o \sb s  \B v\sb s
 (\B k, t) + \B f\sb n(\B k, t), 
  \\    \G \sb n=\g\sb n + \frac{\alpha \r\sb s}{\r \sb n}  \bar\o \sb s  + \nu\sb n k^2\,,\!\!\!\!\!  && ~\G \sb s=\g \sb s + \a \,
  \bar\o \sb s+  \nu'\sb s k^2\ .
\end{eqnarray}\end{subequations}
Multiplying \Eqs{NSE2a} and \eq{NSE2b} by $\B v\sb s$, and  $\B v\sb n$,  respectively  and averaging, one gets equations for velocity  correlations
$G\sb{nn}$, $G\sb{ss}$ and cross-correlation $G\sb{sn}$, defined by \Eqs{def6}:
\begin{subequations}\label{CF}\begin{eqnarray}\label{CFa} 
  \Big[\frac{\p  }{\p t}+ 2 \G \sb s \Big ] G\sb {ss}   &=& 2 \alpha\,  \bar \o \sb s  G\sb {sn}  + 2 \mbox{Re} \big[ \Phi\sb
  {ss}\big ]\,,  ~~~~~~~~
 \\ \label{NSE2b}
 \Big[\frac{\p  }{\p t}+ 2 \G \sb n \Big ] G\sb {nn}   &=& 2 \frac{\a \r \sb s}{\r \sb n  } \bar \o \sb s  G\sb {sn}  + 2 \mbox{Re}
 \big[ \Phi\sb {nn}\big ]\,,  ~~~~~~~~
 \\
 \label{NSE2c}
 \Big[\frac{\p  }{\p t}+ \G \sb s + \G \sb n \Big ] G\sb {sn}   &=& \bar\o \sb s\Big[\frac{\a \r \sb s}{\r \sb n  }   G\sb {nn} +
 \alpha \sb s   G\sb {ss}\Big]+ \br &&  ~\mbox{Re} \big[ \Phi\sb {sn}+ \Phi\sb {ns}\big ]\ .  ~~~~~~~~
 \end{eqnarray}\end{subequations}
These equations involve  yet unknown cross-correlations   of the velocities and the forces, $\Phi_{\dots}$,  defined similarly to
\Eqs{corr}: 
\begin{subequations}\label{corr1}
  \begin{eqnarray}\label{corr-nn}
    \< \B f\sb n(\B k,t)\*\B v^* \sb n(\B k',t)  \>&=& (2\pi)^3 \Phi\sb{nn}(\B k)\, \d(\B k-\B k')\,, \\ 
    \label{corr-ss}
    \< \B f\sb s(\B k,t)\*\B v^* \sb s(\B k',t)  \>&=& (2\pi)^3 \Phi\sb{ss}(\B k)\, \d(\B k-\B k')\,, \\ 
    \label{corr-ns}
    \< \B f\sb n(\B k,t)\*\B v^* \sb s(\B k',t)  \>&=& (2\pi)^3 \Phi\sb{ns}(\B k)\, \d(\B k-\B k')\,, \\ 
    \< \B f\sb s(\B k,t)\*\B v^* \sb n(\B k',t)  \>&=& (2\pi)^3 \Phi\sb{sn}(\B k)\, \d(\B k-\B k')\ .~~~~~
\end{eqnarray}\end{subequations}
To find these correlations, we rewrite  \Eqs{NSE2} in Fourier $\o$-representation:
  \begin{subequations}\label{NSE3}\begin{eqnarray}\label{NSE3a} 
 \big [i\o  + \G \sb s  \big ] \~ {\B v} \sb s(\B k, \o) &=&\alpha \sb s \bar \o \sb s \~{\B v}\sb n(\B k, \o)  + \~{\B f}\sb s(\B k,
 \o)\,,  ~~~~~~~~
 \\ \label{NSE3b}
  \big [i\o + \G \sb n  \big ]  \~{\B v}\sb n(\B k, \o)  & =& \frac{\a \r \sb s}{\r \sb n} \bar \o \sb s  \~ {\B v}\sb s (\B k, \o) +
  \~{\B f} \sb n(\B k, \o) \,,
 \end{eqnarray}\end{subequations}
were $\~ {\B v}\sb{\dots}$ and  $\~ {\B f}\sb{\dots}$ denote Fourier transforms of the corresponding functions. The solution of
linear \Eqs{NSE3} reads:
\begin{subequations}\label{sol3}\begin{eqnarray}\label{sol3a}
\~ {\B v} \sb s  &=& -  \big[ (i\o  + \G \sb n) \~ {\B f}\sb s    + \alpha \sb s \bar\o \sb s  \~ {\B f}\sb n \big ]/ \D  \,, \\
\~ {\B v} \sb n  &=& -  \big[ (i\o  + \G \sb s) \~ {\B f}\sb n   + \alpha \sb s \bar \o \sb s  \~ {\B f}\sb s \big ]/ \D  \,,
\end{eqnarray}
\begin{eqnarray}
\D &\=& \o^2 + \frac{\a^2 \r \sb s}{\r \sb n}   \bar\o \sb s^2 - \G \sb s \G \sb n - i \o ( \G \sb s + \G \sb n)\,,  ~~
 \end{eqnarray}\end{subequations}
where for brevity we skipped the arguments $(\B k, \o)$ in all functions.   Multiplying the two \Eqs{sol3} by $\~{\B f} \sb n$ and $\~{\B f}
\sb s$,  respectively  and averaging, we get four equations for (cross)-correlations   $\~ \Phi_{\dots}(\B k,\o)$  in $\omega$-representations via two
correlations $f^2\sb {ss} (\B k)$ and $f^2\sb {nn} (\B k)$ of random forces, \Eqs{fdef}. By integration of the results over $\o$ one
may, in principle, get the simultaneous cross-correlation functions $  \Phi\sb {nn}(\B k)$, $  \Phi\sb {ss}(\B k)$, $  \Phi\sb {sn}(\B
k)$ and  $  \Phi\sb {ns}(\B k)$, expressed via $f^2\sb {ss} (\B k)$ and $f^2\sb {nn} (\B k)$. However, it suffices to realize that
$\Phi\sb {sn}(\B k) = \Phi\sb {ns}(\B k) = 0$. This instantly simplifies the stationary ($\p /\p t=0$) \Eq{NSE2c}  to
\begin{equation}
    \Big[\G \sb s + \G \sb n \Big ] G\sb {sn} = \o \sb s\Big[\frac{\a \r \sb s}{\r \sb n}   G\sb {nn} + \alpha \sb s   G\sb
    {ss}\Big]\,,
\end{equation}
and we get a relationship between $G\sb{nn}(\B k)$,  $G\sb{ss}(\B k)$ and $G\sb{sn}(\B k)$, that is equivalent [with account of
\Eqs{def1a}] to   \Eq{lim1}.

\end{document}